\documentclass[11pt,a4paper]{article}
 
\usepackage{amsmath,amsthm,amssymb}
\usepackage{graphics,graphicx}
\usepackage{epsfig}
\usepackage{multicol}
\usepackage{color}
\makeatletter
\@addtoreset{equation}{section}
\makeatother

\usepackage{braket}
\usepackage{mathtools}
\usepackage{amssymb}

\begin{document}

%Title%%%%%%%%%%%%%%%%%%%%%%%%%%%%%%%%%%%%%%%%%%%%%%%%%%%%%%%%%%%%%%%%%%
%%%%%%%%%%%%%%%%%%%%%%%%%%%%%%%%%%%%%%%%%%%%%%%%%%%%%%%%%%%%%%%%%%%%%
\begin{center}
\LARGE{\bf How does the Planck scale affect qubits?}
\end{center}

\begin{center}
\large{\bf Matthew J. Lake} ${}^{a,b}$\footnote{matthew5@mail.sysu.edu.cn}
\end{center}
\begin{center}
\emph{ ${}^a$ School of Physics, Sun Yat-Sen University, \\ Guangzhou 510275, People's Republic of China \\ }
\emph{ ${}^b$ Frankfurt Institute for Advanced Studies (FIAS), \\ Ruth-Moufang-Str. 1, D-60438, Frankfurt am Main, Germany}
\vspace{0.1cm}
\end{center}

%Abstract%%%%%%%%%%%%%%%%%%%%%%%%%%%%%%%%%%%%%%%%%%%%%%%%%%%%%%%%%%%%%%%%
%%%%%%%%%%%%%%%%%%%%%%%%%%%%%%%%%%%%%%%%%%%%%%%%%%%%%%%%%%%%%%%%%%%%%

\begin{abstract}
Gedanken experiments in quantum gravity motivate generalised uncertainty relations (GURs) implying deviations from the standard quantum statistics close to the Planck scale. 
These deviations have been extensively investigated for the non-spin part of the wave function but existing models tacitly assume that spin states remain unaffected by the quantisation of the background in which the quantum matter propagates. 
Here, we explore a new model of nonlocal geometry in which the Planck-scale smearing of classical points generates GURs for angular momentum. 
These, in turn, imply an analogous generalisation of the spin uncertainty relations. 
The new relations correspond to a novel representation of {\rm SU(2)} that acts nontrivially on both subspaces of the composite state describing matter-geometry interactions. 
For single particles each spin matrix has four independent eigenvectors, corresponding to two $2$-fold degenerate eigenvalues $\pm (\hbar + \beta)/2$, where $\beta$ is a small correction to the effective Planck's constant. 
These represent the spin states of a quantum particle immersed in a quantum background geometry and the correction by $\beta$ emerges as a direct result of the interaction terms. 
In addition to the canonical qubits states, $\ket{0} = \ket{\uparrow}$ and $\ket{1} = \ket{\downarrow}$, there exist two new eigenstates in which the spin of the particle becomes entangled with the spin sector of the fluctuating spacetime. 
We explore ways to empirically distinguish the resulting `geometric' qubits, $\ket{0'}$ and $\ket{1'}$, from their canonical counterparts.
\end{abstract}

%Keywords
{\bf Keywords}: generalised uncertainty relations; nonlocal geometry; quantum information \\ theory; quantum reference frames; Planck scale; de Sitter scale; dark energy

\tableofcontents

%%%%%%%%%%%%%%%%%%%%%%%%%%%%%%%%%%%%%%%%%%%%%%%%%%%%%%%%%%
%%%%%%%%%%%%%%%%%%%%%%%%%%%%%%%%%%%%%%%%%%%%%%%%%%%%%%%%%%
\section{Introduction} \label{Sec.1}

The Heisenberg uncertainty principle (HUP) contains the essence of quantum theory. 
It can be motivated heuristically by the Heisenberg microscope thought experiment, %\cite{Rae2002},
giving
\begin{eqnarray} \label{HUP-1}
\Delta x^i \, \Delta p_j \gtrsim \frac{\hbar}{2} \delta^{i}{}_{j} \, ,
\end{eqnarray} 
or derived rigorously from the quantum formalism. %\cite{Isham1995}.
In the latter it follows from the Schr{\" o}dinger-Robertson inequality $\Delta_\psi O_1 \, \Delta_\psi O_2 \geq (1/2) |\braket{[\hat{O}_1,\hat{O}_2]}_{\psi}|$, where $\Delta_\psi O = \sqrt{\braket{\hat{O}^2}_{\psi} - \braket{\hat{O}}_{\psi}^2}$, plus the canonical position-momentum commutator,
\begin{eqnarray} \label{[x,p]-1}
[\hat{x}^i,\hat{p}_j] = i\hbar\delta^{i}{}_{j} \, \mathbb{I} \, ,
\end{eqnarray}
yielding
\begin{eqnarray} \label{HUP-2}
\Delta_\psi x^i \, \Delta_\psi p_j  \geq \frac{\hbar}{2} \delta^{i}{}_{j} \, .
\end{eqnarray}
In this case the inequality is precise and the heuristic uncertainties, $\Delta x^i$ and $\Delta p_j$, are replaced by well defined standard deviations, $\Delta_\psi x^i$ and $\Delta_\psi p_j$. 
The HUP (\ref {HUP-2}) is a fundamental consequence of wave-particle duality and follows directly from the properties of the Fourier transform, plus the canonical de Broglie relation $\bold{p} = \hbar\bold{k}$. 

In recent years, the Heisenberg microscope argument as been extended to include the effects of the gravitational attraction between the massive particle and the probing photon \cite{Adler:1999bu,Scardigli:1999jh}. 
This motivates the generalised uncertainty principle (GUP), 
\begin{eqnarray} \label{GUP-1}
\Delta x^i \gtrsim \frac{\hbar}{2\Delta p_j} \delta^{i}{}_{j} \left[1 + \alpha_0 \frac{2G}{\hbar c^3}(\Delta p_j)^2\right] \, ,
\end{eqnarray} 
where $\alpha_0 \sim \mathcal{O}(1)$ is a numerical constant of order unity. 
Conversely, neglecting gravitational attraction but incorporating repulsive effects due to the presence of a background dark energy density, $\rho_{\Lambda} = \Lambda c^2/(8\pi G)$, where $\Lambda \simeq 10^{-56} \, {\rm cm}^{-2}$ is the cosmological constant \cite{Hobson:2006se}, motivates the extended uncertainty principle (EUP). 
This may be written as 
\begin{eqnarray} \label{EUP-1}
\Delta p_j \gtrsim \frac{\hbar}{2\Delta x^i} \delta^{i}{}_{j} \left[1 + 2\eta_0 \Lambda (\Delta x^i)^2\right] \, , 
\end{eqnarray} 
where $\eta_0 \sim \mathcal{O}(1)$ is a numerical constant \cite{Bolen:2004sq,Park:2007az,Bambi:2007ty}. 
The GUP (\ref{GUP-1}) implies the existence of a minimum length scale, of the order of the Planck length, whereas the EUP (\ref{EUP-1}) implies a minimum momentum scale, of the order of the de Sitter momentum $\sim (\hbar/c)\sqrt{\Lambda}$. 

For later convenience, we define the Planck length and mass scales as 
\begin{eqnarray} \label{Planck_scales}
l_{\rm Pl} = \sqrt{\frac{\hbar G}{c^3}} \simeq 10^{-33} \, {\rm cm} \, , \quad m_{\rm Pl} = \sqrt{\frac{\hbar c}{G}} \simeq 10^{-5} \, {\rm g} \, ,
\end{eqnarray}
and the de Sitter length and mass scales as 
\begin{eqnarray} \label{dS_scales}
l_{\rm dS} = \sqrt{\frac{3}{\Lambda}} \simeq 10^{28} \, {\rm cm} \, , \quad m_{\rm dS} = \frac{\hbar}{c}\sqrt{\frac{\Lambda}{3}} \simeq 10^{-66} \, {\rm g} \, ,
\end{eqnarray} 
respectively. 
Incorporating the effects of both gravitational attraction and dark energy repulsion motivates the extended generalised uncertainty principle (EGUP), 
\begin{eqnarray} \label{EGUP-1}
\Delta x^i\Delta p_j \gtrsim \frac{\hbar}{2} \delta^{i}{}_{j} \left[1 + \alpha_0 \frac{2G}{\hbar c^3}(\Delta p_j)^2 + 2\eta_0\Lambda (\Delta x^i)^2\right] \, .
%\nonumber\\
\end{eqnarray}
This implies the existence of both a minimum length, of order $l_{\rm Pl}$, and a minimum momentum, of order $m_{\rm dS}c$ \cite{Bolen:2004sq,Park:2007az,Bambi:2007ty}. 

Various attempts have been made to derive quantum gravity-inspired GURs from a modified quantum formalism but traditional models based on modified commutation relations face severe difficulties \cite{Kempf:1996ss}. 
The four most serious problems are (i) violation of the equivalence principle, (ii) reference frame-dependence of the minimum length, (iii) violation of Lorentz invariance in the relativistic limit, and (iv) the inability to construct sensible multiparticle states, known as the soccer ball problem \cite{Hossenfelder:2012jw,Tawfik:2014zca}.

Recently a new model was proposed, which successfully generates the GUP, EUP and EGUP, but which solves (or rather, evades) problems (i)-(iv) \cite{Lake:2018zeg,Lake:2019nmn,Lake:2020chb,Lake:2020rwc}. 
In this paper, we extrapolate its consequences for orbital angular momentum and spin, focussing on the structure of the modified spin-measurement operators, and their associated eigenstates, for one- and two-particle systems. 
Our analysis shows that, although the predicted deviations from canonical quantum statistics remain small, and probably unmeasurable with current experimental technology, the algebraic and geometric structure of the theory is rich and highly novel. 
Furthermore, the theory is immediately relevant for studies of quantum information, since, although the canonical qubits $\ket{0}$ and $\ket{1}$ are purely abstract, their most obvious realisation in real world quantum systems is as spin states, e.g., 
\begin{eqnarray} \label{qubits}
|0\rangle = \ket{\uparrow_{\rm z}} \, , \quad |1\rangle = \ket{\downarrow_{\rm z}} \, .
\end{eqnarray}

We demonstrate that, in the EGUP-compatible spin model derived in \cite{Lake:2019nmn,Lake:2020rwc}, the canonical qubits (\ref{qubits}) are generalised such that $\left\{|0\rangle,|1\rangle \right\} \rightarrow \left\{|0\rangle,|1\rangle;|0'\rangle,|1'\rangle\right\}$, where $|0'\rangle$ and $|1'\rangle$ represent spins that are entangled with the background geometry. 
These `geometric' qubits are empirically indistinguishable from their canonical counterparts, via simultaneous measurements of $\hat{S}_{\rm z}$ and $\hat{S}^2$, despite their completely different geometric structure. 
Nonetheless, they may in principle be probed by an appropriately subtle measurement technique, capable of exploiting their entanglement with the background space \cite{Lake:2019nmn,Lake:2020rwc}. 

The structure of this paper is as follows. 
In Sec. \ref{Sec.2} we review the basic formalism of the smeared space model, derive the GUP, EUP and EGUP, and introduce GURs for orbital angular momentum. 
Analogous GURs for spin measurements are introduced in Sec. \ref{Sec.3} and the algebraic structure of the generalised spin operators is analysed. 
We first analyse one-particle states in Sec. \ref{Sec.3.1}, before considering two-particle systems in Sec. \ref{Sec.3.2}. 
Special attention is payed to the generalisation of canonical Bell states, which are dealt with in Sec. \ref{Sec.3.3}. 
In Sec. \ref{Sec.4}, we derive the explicit form of the four-dimensional representation of ${\rm SU(2)}$ that acts on the composite state describing matter-geometry interactions. 
Sec. \ref{Sec.5} contains a summary of our conclusions and a discussion of prospects for future work. 
For convenience, a brief recap of the treatment of one- and two-particle systems in canonical quantum mechanics is given in Appendix \ref{Appendix-A}. 
This contains only well known textbook material, which is included as a reference, to which the reader can compare the generalised theory presented in Sec. \ref{Sec.3}. 
For this reason, the structures of Secs. (\ref{Sec.3.1})-(\ref{Sec.3.3}) and Secs. (\ref{Sec.A.1})-(\ref{Sec.A.3}) are analogous, so that the relevant results may be easily contrasted. 
In order to distinguish between analogous operators in the canonical and smeared space theories, we use lower case letters to denote former and upper case letters for the latter, e.g., $\hat{S}_{\rm z}$ and $\hat{S}^2$ as opposed to $\hat{s}_{\rm z}$ and $\hat{s}^2$. 
Due to their relatively large size, the explicit forms of $\hat{S}_{\rm z}$, $\hat{S}^2$ and $\hat{S}_{\pm}$ for the smeared two-particle state are given in Appendix \ref{Appendix-B}.

%%%%%%%%%%%%%%%%%%%%%%%%%%%%%%%%%%%%%%%%%%%%%%%%%%%%%%%%%%
%%%%%%%%%%%%%%%%%%%%%%%%%%%%%%%%%%%%%%%%%%%%%%%%%%%%%%%%%%
\section{The smeared space model of nonlocal geometry} \label{Sec.2}

The model proposed in \cite{Lake:2018zeg,Lake:2019nmn} is based on the modified de Broglie relation
\begin{eqnarray} \label{mod_dB*}
\bold{p}' = \hbar\bold{k} + \beta(\bold{k}'-\bold{k}) \, ,
\end{eqnarray}
where $\beta$ is a very small constant of action that is determined by the ratio of the Planck and dark energy densities, 
\begin{eqnarray} \label{beta}
\beta = 2\hbar \sqrt{\frac{\rho_{\Lambda}}{\rho_{\rm Pl}}} \simeq \hbar \times 10^{-61} \, . 
\end{eqnarray}
Here, $\bold{p}'$ represents the observable momentum of a quantum particle propagating in a quantum background space and the non-canonical term in Eq. (\ref{mod_dB*}) describes additional momentum `kicks' due to fluctuations of the geometry. 

In this model, $\bold{p} = \hbar\bold{k}$ and $\bold{p}' - \bold{p} = \beta(\bold{k}'-\bold{k})$ are independent degrees of freedom and the non-spin part of the quantum state of the composite system describing matter-geometry interactions can be written as
\begin{eqnarray} \label{|Psi>}
|\Psi\rangle = |\psi\rangle \otimes |g\rangle \, ,
\end{eqnarray}
where $|\psi\rangle = \int \psi(\bold{x})|\bold{x}\rangle{\rm d}^3{\rm x} = \int \tilde{\psi}_{\hbar}(\bold{p})|\bold{p}\rangle{\rm d}^3{\rm p}$. 
The additional ket $|g\rangle$ is given by
\begin{eqnarray} \label{|gi>}
|g\rangle = \int g(\bold{x}'-\bold{x}) |\bold{x}'-\bold{x}\rangle{\rm d}^3{\rm x}' 
= \int \tilde{g}_{\beta}(\bold{p}'-\bold{p}) |\bold{p}'-\bold{p}\rangle{\rm d}^3{\rm p}' \, ,  
\end{eqnarray}
where $\langle g|g \rangle = \int |g(\bold{x}'-\bold{x})|^2 {\rm d}^{3}{\rm x}' = \int | \tilde{g}_{\beta}(\bold{p}'-\bold{p})|^2 {\rm d}^{3}{\rm p}' = 1$. 
This describes the quantum state of the background geometry, or, more specifically, the {\it influence} that quantum fluctuations of the background geometry have on the canonical quantum particle described by $|\psi\rangle$.

For simplicity, $g(\bold{x}'-\bold{x})$ can be thought of as a Gaussian distribution with standard deviation $\sigma_g \simeq l_{\rm Pl}$. 
Its momentum space representation $\tilde{g}_{\beta}(\bold{p}'-\bold{p})$ is given by the Fourier transform, where the transformation is performed at the scale $\beta$ rather than $\hbar$ \cite{Lake:2019nmn,Lake:2020rwc}. 
This yields the standard deviation $\tilde{\sigma}_g = \beta/(2\sigma_g) \simeq m_{\rm dS}c$. 
In this way, the model based on Eq. (\ref{mod_dB*}) represents a form a nonlocal geometry in which points in the canonical quantum phase space become smeared over finite-width volumes of order $\sim l_{\rm Pl}^3$ and $\sim (m_{\rm dS}c)^3$, in each representation of the theory \cite{Lake:2020rwc}. 
The smearing functions $g(\bold{x}'-\bold{x})$ and $\tilde{g}_{\beta}(\bold{p}'-\bold{p})$ are interpreted as quantum probability amplitudes for the coherent transitions $\bold{x} \leftrightarrow \bold{x}'$ and $\bold{p} \leftrightarrow \bold{p}'$, respectively, and the primed variables represent the possible measured values of the particle's position and momentum, incorporating its interaction with the fluctuating quantum background. 
%\textcolor{blue}{It is assumed that the transitions $\bold{x} \leftrightarrow \bold{x}'$ and $\bold{p} \leftrightarrow \bold{p}'$ also exchange the canonical amplitudes, such that $\psi(\bold{x}) \leftrightarrow \psi(\bold{x}')$ and $\tilde{\psi}_{\hbar}(\bold{p}) \leftrightarrow \tilde{\psi}_{\hbar}(\bold{p}')$, respectively.}

Since an individual measurement of $\bold{x}'$ cannot determine which of the canonical phase space point(s) $\bold{x}$ underwent the transition $\bold{x} \leftrightarrow \bold{x}'$, we must sum over all possible values, and analogous reasoning holds in the momentum space picture. 
For Gaussian smearing, $\bold{x}' = \bold{x}$ and $\bold{p}' = \bold{p}$ are the most probable measurement outcomes, but fluctuations within one standard deviation of $|g(\bold{x}' - \bold{x})|^2$ or $|\tilde{g}_{\beta}(\bold{p}' - \bold{p})|^2$ remain relatively likely \cite{Lake:2018zeg,Lake:2019nmn,Lake:2020chb,Lake:2020rwc}. 
The associated probability distributions take the form of convolutions, i.e., ${\rm d}P(\bold{x}'|\Psi) = (|\psi|^2*|g|^2)(\bold{x}'){\rm d}^{3}{\rm x'}$ and ${\rm d}P(\bold{p}'|\tilde{\Psi}) = (|\tilde{\psi}_{\hbar}|^2*|\tilde{g}_{\beta}|^2)(\bold{p}'){\rm d}^{3}{\rm p'}$, where  
\begin{eqnarray} \label{Psi}
\Psi(\bold{x},\bold{x}') = \psi(\bold{x}) \, g(\bold{x}' - \bold{x}) \, ,  
\end{eqnarray}
\begin{eqnarray} \label{Psi_tilde}
\tilde{\Psi}(\bold{p},\bold{p}') = \tilde{\psi}_{\hbar}(\bold{p}) \, \tilde{g}_{\beta}(\bold{p}' - \bold{p}) \, , 
\end{eqnarray}
are the position space and momentum space wave functions of the interacting matter-plus-geometry states, respectively. 
These give rise to GURs of the form
\begin{eqnarray} \label{smeared-GUP-1}
\Delta_{\Psi}x'^{i} = \sqrt{(\Delta_{\psi}x'^{i})^2 + \sigma_g^2} \, , 
\end{eqnarray}
\begin{eqnarray} \label{smeared-EUP-1}
\Delta_{\Psi}p'_{j} = \sqrt{(\Delta_{\psi}p'_{j})^2 + \tilde{\sigma}_g^2} \, . 
\end{eqnarray}
(Note the difference between $\Psi = \psi g$ and $\psi$ and in the subscripts.)

Using $\sigma_g \simeq l_{\rm Pl}$ and $\tilde{\sigma}_g \simeq m_{\rm dS}c$, substituting for $\Delta_{\psi}x'^{i}$ and $\Delta_{\psi}p'_{j}$ from the HUP (\ref{HUP-2}), and Taylor expanding the right-hand sides of Eqs. (\ref{smeared-GUP-1})-(\ref{smeared-EUP-1}) to first order yields the GUP (\ref{GUP-1}) and EUP (\ref{EUP-1}), respectively, but with the heuristic uncertainties replaced by the standard deviations of well defined probability distributions. 
 
The smeared space GURs can also be rewritten in terms of the generalised position and momentum operators,
\begin{eqnarray} \label{ops_split_Q}
\hat{X}^{i} = \hat{Q}^{i} + \hat{Q}'^{i} = (\hat{q}^{i} \otimes \mathbb{I}) + (\mathbb{I} \otimes \hat{q}'^{i}) \, , 
\end{eqnarray}
\begin{eqnarray} \label{ops_split_Pi}
\hat{P}_{j} = \hat{\Pi}_{j} + \hat{\Pi}'_{j} = (\hat{\pi}_{j} \otimes \mathbb{I}) + (\mathbb{I} \otimes \hat{\pi}'_{j}) \, , 
\end{eqnarray}
where
\begin{eqnarray} \label{QQ'}
\hat{q}^{i} = \int q^{i} \ket{\bold{q}}\bra{\bold{q}} {\rm d}^3{\rm q}  \, , \quad \hat{q}'^{i} = \int q'^{i} \ket{\bold{q}'}\bra{\bold{q}'} {\rm d}^3{\rm q}'  \, , 
\end{eqnarray}
\begin{eqnarray} \label{PiPi'}
\hat{\pi}_{j} = \int \pi_{j} \ket{\boldsymbol{\pi}}\bra{\boldsymbol{\pi}} {\rm d}^3{\rm \pi}  \, , \quad \hat{\pi}'_{j} = \int \pi'_{j} \ket{\boldsymbol{\pi}'}\bra{\boldsymbol{\pi}'} {\rm d}^3{\rm \pi}'  \, , 
\end{eqnarray}
and we have defined the new variables
\begin{eqnarray} \label{new_variables-1}
\bold{q} = \bold{x} \, , \quad \bold{q}' = \bold{x}' - \bold{x} \, , \quad (\bold{X} = \bold{q} + \bold{q}' =  \bold{x}') \, ,
\end{eqnarray}
\begin{eqnarray} \label{new_variables-2}
\boldsymbol{\pi} = \bold{p} \, , \quad \boldsymbol{\pi}' = \bold{p}' - \bold{p} \, , \quad (\bold{P} = \boldsymbol{\pi} + \boldsymbol{\pi}' = \bold{p}') \, .
\end{eqnarray}
Equations (\ref{smeared-GUP-1})-(\ref{smeared-EUP-1}) then take the form
\begin{eqnarray} \label{comp-1}
\Delta_{\Psi}X^{i} &=& \sqrt{(\Delta_{\Psi}Q^{i})^2 + (\Delta_{\Psi}Q'^{i})^2} \, ,
\end{eqnarray}
\begin{eqnarray} \label{comp-2}
\Delta_{\Psi}P_{j} &=& \sqrt{(\Delta_{\Psi}\Pi_{j})^2 + (\Delta_{\Psi}\Pi'_{j})^2} \, , 
\end{eqnarray}
since $\Delta_{\Psi}Q^{i} = \Delta_{\psi}x'^{i}$, $\Delta_{\Psi}Q'^{i} = \Delta_{g}x'^{i} \equiv \sigma_g$ and $\Delta_{\Psi}\Pi_{j} = \Delta_{\psi}p'_{j}$, $\Delta_{\Psi}\Pi'_{j}  = \Delta_g p'_{j} \equiv \tilde{\sigma}_g$, yielding $\Delta_{\Psi}X^{i} = \Delta_{\Psi}x'^{i}$ and $\Delta_{\Psi}P_{j} = \Delta_{\Psi}p'_{j}$, respectively \cite{Lake:2019nmn,Lake:2020rwc}. 
Combining Eqs. (\ref{comp-1}) and (\ref{comp-2}), Taylor expanding to first order and ignoring subdominant terms, and again using $\sigma_g \simeq l_{\rm Pl}$ and $\tilde{\sigma}_g \simeq m_{\rm dS}c$, yields
\begin{eqnarray} \label{EGUP-2}
\Delta_{\Psi}X^{i} \, \Delta_{\Psi}P_{j} \gtrsim \frac{\hbar}{2}\delta^{i}{}_{j} \, [1 + \alpha (\Delta_{\Psi}X^{i})^2 + \eta (\Delta_{\Psi}P_{i})^2] \, ,
\end{eqnarray}
with $\alpha = 4G/(\hbar c^3)$ and $\eta = \Lambda/6$. 
This is the EGUP (\ref{EGUP-1}), expressed in terms of the {\it observable} uncertainties $\Delta_{\Psi}X^{i}$ and $\Delta_{\Psi}P_{j}$, which are derived from the smeared-space joint probability distribution $|\Psi|^2$. 

To obtain Eqs. (\ref{comp-1})-(\ref{comp-2}), directly from Eqs. (\ref{ops_split_Q})-(\ref{ops_split_Pi}), we made use of the fact that ${\rm cov}_{\Psi}(\hat{Q}^{i},\hat{Q}'^{i}) = {\rm cov}_{\Psi}(\hat{Q}'^{i},\hat{Q}^{i}) = 0$ and 
${\rm cov}_{\Psi}(\hat{\Pi}_{j},\hat{\Pi}'_{j}) = {\rm cov}_{\Psi}(\hat{\Pi}'_{j},\hat{\Pi}_{j}) = 0$, where ${\rm cov}(X,Y) = \braket{XY} - \braket{X}\braket{Y}$ is the covariance of the random variables $X$ and $Y$. 
The operator pairs $\hat{Q}^{i}$, $\hat{Q}'^{i}$ and $\hat{\Pi}_{j}$, $\hat{\Pi}'_{j}$ are uncorrelated because they act on disjoint subspaces of the total state $\ket{\Psi}$ (\ref{|Psi>}). 
In this basis, all unprimed operators commute with all primed operators, and the subcomponents of $\hat{X}^{i}$ and $\hat{P}_{j}$ satisfy the algebra
\begin{subequations}
\begin{eqnarray} \label{[Q,Pi]}
[\hat{Q}^{i},\hat{\Pi}_{j}] = i\hbar \delta^{i}{}_{j} \, \mathbb{I} \, , \quad [\hat{Q}'^{i},\hat{\Pi}'_{j}]  = i\beta \delta^{i}{}_{j} \, \mathbb{I} \, ,
\end{eqnarray}
\begin{eqnarray} \label{mixed_comm-1}
[\hat{Q}^{i},\hat{\Pi}'_{j}] = [\hat{Q}'^{i},\hat{\Pi}_{j}]  = 0 \, ,
\end{eqnarray}
\begin{eqnarray} \label{QQ_commutators}
[\hat{Q}^{i},\hat{Q}^{j}] = [\hat{Q}'^{i},\hat{Q}'^{j}] = 0 \, ,
\end{eqnarray}
\begin{eqnarray} \label{PiPi_commutators}
\quad [\hat{\Pi}_{i}, \hat{\Pi}_{j}] = [\hat{\Pi}'_{i}, \hat{\Pi}'_{j}] = 0 \, ,
\end{eqnarray}
\begin{eqnarray} \label{remaining_commutators}
[\hat{Q}^{i},\hat{Q}'^{j}] = 0 \, , \quad [\hat{\Pi}_{i}, \hat{\Pi}'_{j}] = 0 \, ,
\end{eqnarray}
\end{subequations}
whose individual relations combine to give
\begin{subequations}
\begin{eqnarray} \label{[X,P]}
[\hat{X}^{i},\hat{P}_{j}] = i(\hbar + \beta)\delta^{i}{}_{j} \, \mathbb{I} \, ,
\end{eqnarray}
\begin{eqnarray} \label{XX_PP_commutators}
[\hat{X}^{i},\hat{X}^{j}] = 0 \, , \quad [\hat{P}_{i},\hat{P}_{j}] = 0 \, .
\end{eqnarray}
\end{subequations}
Together, (\ref{[Q,Pi]}) and (\ref{mixed_comm-1}) yield Eq. (\ref{[X,P]}) while (\ref{QQ_commutators})-(\ref{remaining_commutators}) yield Eqs. (\ref{XX_PP_commutators}). 

The commutation relations (\ref{[X,P]})-(\ref{XX_PP_commutators}) are simply a rescaled version of the canonical Heisenberg algebra, with $\hbar \rightarrow \hbar + \beta$, but the most general uncertainty relation for smeared position and momentum measurements is not of the canonical Heisenberg type. 
It takes the form
\begin{eqnarray} \label{smeared_XP_UR}
\Delta_{\Psi}X^{i} \, \Delta_{\Psi}P_{j} \geq . \, . \, . \, \geq \left(\frac{\hbar + \beta}{2}\right) \delta^{i}{}_{j} \, \mathbb{I} \, ,
\end{eqnarray}
where the dots in the middle represent a sum of terms that is generically greater than or equal to the Schr{\" o}dinger-Robertson bound on the far right-hand side. 
The GUP, EUP and EGUP all arise as different limits of this general relation \cite{Lake:2018zeg,Lake:2019nmn,Lake:2020chb,Lake:2020rwc}. 

In other words, the simplicity of Eq. (\ref{[X,P]}) is deceptive. 
At first sight, one may be forgiven for thinking that the presence of $(\hbar + \beta)$ rather than $\hbar$ represents a rescaling of our measurement units, without introducing new physics, but the subalgebra structure (\ref{[Q,Pi]})-(\ref{remaining_commutators}) shows that this is not the case. 
Ultimately, it this subalgebra that is responsible for the rescaling $\hbar \rightarrow \hbar + \beta$ {\it and} the smeared-space GUR (\ref{smeared_XP_UR}), from which the GUP, EUP and EGUP can be recovered \cite{Lake:2018zeg,Lake:2020rwc}. 
This may be seen, explicitly, by considering the position space representation of the wave mechanics picture, in which
\begin{eqnarray} \label{X^i_wave_mech}
\hat{X}^{i} = x^{i} + (x'^{i} - x^{i}) = x'^{i} \, .
\end{eqnarray}
The operator $\hat{{\rm d}}_{j} = i(\hbar+\beta)^{-1}\hat{P}_{j}$ then generates infinitesimal shifts in the $x'^{j}$ direction. 
However, it is important to note that $\hat{P}_{j} \neq -i(\hbar + \beta)\partial/\partial x'^{j}$. 
Instead, the correct expression for $\hat{P}_{j}$ is 
\begin{eqnarray} \label{P_j_wave_mech}
\hat{P}_{j} = -i\hbar \frac{\partial}{\partial x^{j}}\Big|_{\bold{x}'-\bold{x} = {\rm const.}} - i\beta \frac{\partial}{\partial (x'^{j}-x^{j})}\Big|_{\bold{x} = {\rm const.}} \, .
\end{eqnarray}
The corresponding shift-isometry generator is parameterised in terms of the dimensionless ratio,
\begin{eqnarray} \label{delta}
\delta = \hbar/\beta \simeq 10^{-61} \, , 
\end{eqnarray}
as
\begin{eqnarray} \label{d_j}
\hat{{\rm d}}_{j}(x') = \frac{1}{1+\delta}\hat{{\rm d}}_{j}(x) + \frac{\delta}{1+\delta}\hat{{\rm d}}_{j}(x'-x) \, . 
\end{eqnarray}
We then have $\exp[i \, \bold{a}.\hat{\bold{P}}/(\hbar+\beta)]\bold{x}' = (\bold{x} + \bold{a}/(1+\delta)) + (\bold{x}' - \bold{x} + \delta\bold{a}/(1+\delta)) = \bold{x}' + \bold{a}$. 
The key point is that each sub-shift is associated with a different probability amplitude. 
The first is associated with $\psi(\bold{x})$ whereas the second represents an additional shift, induced by fluctuations of the background, and is associated with $g(\bold{x}'-\bold{x})$. 
The composite probability amplitude for both shifts is $\Psi(\bold{x},\bold{x}')$ (\ref{Psi}).

In the remainder of this section, we show that similar considerations hold for smeared angular momentum measurements. 
The canonical so(3) Lie algebra is rescaled such that $\hbar \rightarrow \hbar + \beta$, but this does not arise from a change of units. 
Instead, it is the consequence of a complex subalgebra structure that also generates GURs. 

The smeared space angular momentum operators are defined as 
\begin{eqnarray} \label{L_i_subcomponents}
\hat{L}_{i} = \epsilon_{ij}{}^{k}\hat{X}^{j}\hat{P}_{k} \, , 
\end{eqnarray}
where $\epsilon_{ij}{}^{k}$ is the Levi-Civita symbol and $\hat{X}^{j}$, $\hat{P}_{k}$ are given by Eqs. (\ref{ops_split_Q})-(\ref{ops_split_Pi}), respectively. 
In terms of the subcomponents $\left\{\hat{Q}^{i},\hat{Q}'^{i},\hat{\Pi}_{j},\hat{\Pi}'_{j}\right\}$ these may be decomposed as 
\begin{eqnarray} \label{L_sum*}
\hat{L}_{i} = \hat{\mathcal{L}}_{i} + \hat{\mathcal{L}}'_{i} + \hat{\mathbb{L}}_{i} \, , 
\end{eqnarray}
where 
\begin{subequations}
\begin{eqnarray} \label{LL'}
\hat{\mathcal{L}}_{i} = \epsilon_{ij}{}^{k}\hat{Q}^{j}\hat{\Pi}_{k} \, , \quad \hat{\mathcal{L}}'_{i} = \epsilon_{ij}{}^{k}\hat{Q}'^{j}\hat{\Pi}'_{k} \, ,
\end{eqnarray}
\begin{eqnarray} \label{mathbbL}
\hat{\mathbb{L}}_{i} = \epsilon_{ij}{}^{k}(\hat{Q}'^{j}\hat{\Pi}_{k} + \hat{Q}^{j}\hat{\Pi}'_{k}) \, .
\end{eqnarray}
\end{subequations}
The angular momentum subcomponents $\left\{\hat{\mathcal{L}}_{i},\hat{\mathcal{L}}'_{i},\hat{\mathbb{L}}_{i}\right\}$ satisfy the subalgebra
\begin{subequations} \label{rearrange-L*}
%1*
\begin{equation} \label{rearrange-L.1*}
[\hat{\mathcal{L}}_{i},\hat{\mathcal{L}}_{j}] = i\hbar \, \epsilon_{ij}{}^{k} \hat{\mathcal{L}}_{k} \, , \quad [\hat{\mathcal{L}}'_{i},\hat{\mathcal{L}}'_{j}] = i\beta \, \epsilon_{ij}{}^{k} \hat{\mathcal{L}}'_{k} \, , 
\end{equation}
%2*
\begin{equation} \label{rearrange-L.2*}
[\hat{\mathcal{L}}_{i},\hat{\mathcal{L}}'_{j}] = [\hat{\mathcal{L}}'_{i},\hat{\mathcal{L}}_{j}] = 0 \, , 
\end{equation}
%3*
\begin{equation} \label{rearrange-L.3*}
[\hat{\mathcal{L}}_{i},\hat{\mathbb{L}}_{j}] - [\hat{\mathcal{L}}_{j},\hat{\mathbb{L}}_{i}] = i\hbar \, \epsilon_{ij}{}^{k} \hat{\mathbb{L}}_{k} \, , 
\end{equation}
%4*
\begin{equation} \label{rearrange-L.4*}
[\hat{\mathcal{L}}'_{i},\hat{\mathbb{L}}_{j}] - [\hat{\mathcal{L}}'_{j},\hat{\mathbb{L}}_{i}] = i\beta \, \epsilon_{ij}{}^{k} \hat{\mathbb{L}}_{k} \, , 
\end{equation}
%5*
\begin{equation} \label{rearrange-L.5*}
[\hat{\mathbb{L}}_{i},\hat{\mathbb{L}}_{j}] = i\beta \, \epsilon_{ij}{}^{k}\hat{\mathcal{L}}_{k} + i\hbar \, \epsilon_{ij}{}^{m}\hat{\mathcal{L}}'_{m} \, ,
\end{equation}
\end{subequations}
whose individual relations combine to give
\begin{subequations}
\begin{eqnarray} \label{LL_commutator}
[\hat{L}_{i},\hat{L}_{j}] = i(\hbar +\beta) \epsilon_{ij}{}^{k}\hat{L}_{k} \, ,  
\end{eqnarray}
\begin{eqnarray} \label{L^2L_commutator}
[\hat{L}_{i},\hat{L}^2] = 0 \, .
\end{eqnarray}
\end{subequations}

In canonical quantum mechanics the angular momentum commutation relations form a representation of the so(3) Lie algebra, scaled by $\hbar$, yielding uncertainty relations of the form $\Delta_{\psi}l_{i}\Delta_{\psi}l_{j} \geq (\hbar/2) \, |\epsilon_{ij}{}^{k}\braket{\hat{l}_{k}}_{\psi}|$. 
This is because the angular momentum operators are directly identified with the rotation generators, via $\hat{l}_{i} = -i\hbar \, \hat{R}_{i}$, where $\left\{\hat{R}_{i}\right\}_{i=1}^{3}$ perform rotations on the space $\mathbb{R}^3$, spanned by the coordinates $\left\{x^i\right\}_{i=1}^{3}$. 
In the smeared space theory, the identification is more subtle. 
The subcomponents $\hat{\mathcal{L}}_{i} = -i\hbar \, \hat{R}_{i}$ and $\hat{\mathcal{L}}'_{i} = -i\beta \, \hat{R}'_{i}$ generate rotations on the subspaces spanned by $\left\{q^i\right\}_{i=1}^{3}$ and $\left\{q'^i\right\}_{i=1}^{3}$, respectively, but the $\hat{\mathbb{L}}_{i}$ subcomponents do not represent rotations in either domain. 
The $\hat{\mathcal{L}}_{i}$ represent the angular momenta of a material particle, as in the canonical theory, whereas the $\hat{\mathcal{L}}'_{i}$ represent the components of angular momentum carried by the fluctuating background, in which the particle propagates. 
The $\hat{\mathbb{L}}_{i}$ operators determine how the two interact and the associated dimensionless generators are $\hat{\mathbb{R}}_{i} = i \, \hat{\mathbb{L}}_{i}/(\hbar + \beta)$. 
The sum of terms $\hat{\mathcal{R}}_i(q+q') := (1+\delta)^{-1}[\hat{R}_{i}(q) + \delta\hat{R}'_{i}(q')] + \hat{\mathbb{R}}_{i}(q,q')$ then generates rotations on the space spanned by the coordinates $\left\{x'^i\right\}_{i=1}^{3} = \left\{q^i+q'^i\right\}_{i=1}^{3}$. 
We recall that, in the smeared space model, these represent the only {\it observable} values of the particle's position. 

The net result is the rescaled Lie algebra, Eq. (\ref{LL_commutator}), but it is important to realise that $\left\{\hat{L}_{i}\right\}_{i=1}^{3}$ in the smeared space model represent {\it more} than just rotations in classical Euclidean space. 
Instead, they represent rotations in the space spanned by all possible measurement outcomes, $\bold{x}' = \bold{q} + \bold{q}'$. 
This is isomorphic to the classical space, $\mathbb{R}^3$, but the total phase space of the nonlocal geometry is $\mathbb{R}^3 \times \mathbb{R}^3$. 
In \cite{Lake:2018zeg}, this was interpreted as a superposition of Euclidean geometries and the smeared phase space `points' described by $\ket{g}$ were interpreted as quantum reference frames, i.e., as superpositions of classical reference frames \cite{Giacomini:2017zju}.  
This interpretation provides a physical basis for the GUR (\ref{smeared_XP_UR}) and the corresponding uncertainty relations for angular momentum take the form
\begin{eqnarray} \label{L_GUR}
\Delta_{\Psi}L_{i} \, \Delta_{\Psi}L_{j} \geq \dots \geq \left(\frac{\hbar + \beta}{2}\right) |\epsilon_{ij}{}^{k}\braket{\hat{L}_{k}}_{\Psi}| \, ,
\end{eqnarray}
where the sum of terms in the middle is determined by the relation
\begin{eqnarray} \label{DL^2*}
(\Delta_{\Psi}L_{i})^2 &=& (\Delta_{\Psi}\mathcal{L}_{i})^2 + (\Delta_{\Psi}\mathcal{L}'_{i})^2 + (\Delta_{\Psi}\mathbb{L}_{i})^2 
\nonumber\\
&+& {\rm cov}(\hat{\mathcal{L}}_{i},\hat{\mathbb{L}}_{i}) + {\rm cov}(\hat{\mathbb{L}}_{i},\hat{\mathcal{L}}_{i})
\nonumber\\
&+& {\rm cov}(\hat{\mathcal{L}}'_{i},\hat{\mathbb{L}}_{i}) + {\rm cov}(\hat{\mathbb{L}}_{i},\hat{\mathcal{L}}'_{i}) \, .
\end{eqnarray}

%%%%%%%%%%%%%%%%%%%%%%%%%%%%%%%%%%%%%%%%%%%%%%%%%%%%%%%%%%
%%%%%%%%%%%%%%%%%%%%%%%%%%%%%%%%%%%%%%%%%%%%%%%%%%%%%%%%%%
\section{Spin in smeared-space quantum mechanics} \label{Sec.3}

Established physical theories are based on two types of symmetries. 
The first are `external' symmetries, that is, symmetries of the spacetime background in which the matter fields are defined, and the second are `internal' symmetries of the matter Lagrangian, otherwise known as gauge symmetries \cite{Healey:2005zz}. 
In canonical quantum mechanics the angular momentum algebra emerges as a consequence of the rotational invariance of the physical space in which the quantum particle `lives'. 
Since the space is classical, this is {\it identical} to the target space spanned by all possible measured values of the particle's position. 
By contrast, quantum mechanical spin is a consequence of {\rm SU(2)} gauge invariance, where {\rm SU(2)} is the double cover of the classical {\rm SO(3)} rotation group. 

In Sec. \ref{Sec.2} we argued that the canonical shift and rotation isometries must be modified, in a subtle way, when the geometry ceases to be classical and instead becomes nonlocal, in the sense defined by Eq. (\ref{|Psi>}). 
The smearing of spatial points leads to a doubling of the canonical phase space dimensions, $\mathbb{R}^3 \mapsto \mathbb{R}^3 \times \mathbb{R}^3 \cong \mathbb{R}^6$. 
In this scenario, the target space of possible measured values, $\bold{x}'$, is isomorphic, but no longer {\it identical} to the classical physical space, $\mathbb{R}^3$. 
It is nontrivially embedded in the total phase space, $\mathbb{R}^3 \times \mathbb{R}^3$, whose individual subspaces are spanned by $\bold{q} = \bold{x}$ and $\bold{q}' = \bold{x}'-\bold{x}$, respectively. 
At the algebraic level, this is expressed by constructing novel representations of the translation and rotation generators, $\hat{{\rm d}}_j(x')$ and $\hat{{\rm \mathcal{R}}}_i(x')$, that act nontrivially on both the first and second subspaces of the extended phase space. 
We claimed that this split defines the `smearing' of classical symmetries, which is correctly described by the subalgebra structures (\ref{[Q,Pi]})-(\ref{remaining_commutators}) and (\ref{rearrange-L.5*})-(\ref{rearrange-L.5*}). 
These give rise to GURs but preserve the the underlying group structure of canonical quantum mechanics. 
It is then reasonable to ask, if the representations of spacetime symmetries must be modified to account for Planck-scale physics, is this possible without a concomitant `smearing' of gauge symmetries? 

There are strong arguments on both sides. 
On the one hand we may argue that, as internal symmetries, there is no reason why gauge invariances of any kind should be affected by the nonlocality of the background space. 
Preserving the action of the canonical SU(2) representation on the composite state describing matter-geometry interactions, i.e., setting $\hat{s}_i = (\hbar/2) \, \sigma_i \otimes \mathbb{I}$ so that $[\hat{s}_i,\hat{s}_j] = i\hbar \epsilon_{ij}{}^{k}\hat{s}_k$, while smearing the SO(3) representation according to Eqs. (\ref{rearrange-L.5*})-(\ref{rearrange-L.5*}) and (\ref{LL_commutator})-(\ref{L^2L_commutator}), then results in a clear break between the algebraic structures of internal and external angular momentum. 
Furthermore, this break has potentially observable consequences, since smeared symmetry algebras give rise to GURs whereas unsmeared ones do not.  

On the other hand, we may ask if it is possible to construct an alternative representation of the SU(2) generators, which also act nontrivially on both subspaces of the composite state describing matter-geometry interactions, and which describes faithfully the behaviour of quantum mechanical spin in the presence of a nonlocal background. 
We expect the generators of this representation to be split into three subcomponents, by analogy with Eq. (\ref{L_sum*}). 
The first component should act nontrivially only on the first spin subspace of the composite state $\ket{\Psi}_{\rm spin}$, which is associated with canonical quantum matter, whereas the second subcomponent should act nontrivially only on the second subspace, which is associated with spin sector of the background. 
The third subcomponent should act nontrivially on both subspaces and all three subcomponents should have constant matrix elements, as required for generators of {\it internal} symmetries.

In Sec. \ref{Sec.3.1} we construct such a generalised representation and show how it gives rise to spin GURs that are analogous to Eqs. (\ref{L_GUR})-(\ref{DL^2*}) for one-particle states. 
Note, however, that we do not require $\ket{\Psi}_{\rm spin}$ to be separable in any basis, 
\begin{eqnarray} \label{Psi_spin}
\ket{\Psi}_{\rm spin} \neq \ket{\psi}_{\rm spin} \otimes \ket{g}_{\rm spin} \, , 
\end{eqnarray}
since no physical condition compels this. 
As we will see, this is key to the emergence of new spin physics in the smeared space model. 
Following the analysis of one-particle states, two-particle states are dealt with in Sec. \ref{Sec.3.2}, and Bell-type states are explicitly considered in Sec. \ref{Sec.3.3}.   

%%%%%%%%%%%%%%%%%%%%%%%%%%%%%%%%%%%%%%%%%%%%%%%%%%%%%%%%%%
\subsection{One-particle systems in smeared space} \label{Sec.3.1}

We construct the generalised spin-measurement operators, for a single particle propagating in the smeared background, as
\begin{eqnarray} \label{S_sum}
\hat{{\rm S}}_{i} = \hat{\mathcal{S}}_{i} + \hat{\mathcal{S}}'_{i} + \hat{\mathbb{S}}_{i} \, , 
\end{eqnarray}
where
\begin{subequations}
\begin{eqnarray} \label{ss'}
\hat{\mathcal{S}}_{i} = \frac{\hbar}{2} \, \sigma_{i} \otimes \mathbb{I} \, , \quad \hat{\mathcal{S}}'_{i} = \frac{\beta}{2} \, \mathbb{I} \otimes \sigma'_{i} \, , 
\end{eqnarray} 
\begin{eqnarray} \label{mathbb{S}_{i}}
\hat{\mathbb{S}}_{i} = \frac{\sqrt{\hbar\beta}}{2} \, \epsilon_{i}{}^{jk}\sigma_{j} \otimes \sigma'_{k} \, . 
\end{eqnarray}
\end{subequations}

It is straightforward to show that, if both $\sigma_{i}$ and $\sigma'_{i}$ represent spin-1/2 Pauli matrices, the subcomponents satisfy the algebras 
\begin{subequations} \label{rearrange-S}
%1
\begin{equation} \label{rearrange-S.1} 
[\hat{\mathcal{S}}_{i},\hat{\mathcal{S}}_{j}] = i\hbar \, \epsilon_{ij}{}^{k} \hat{\mathcal{S}}_{k} \, ,  \quad [\hat{\mathcal{S}}'_{i},\hat{\mathcal{S}}'_{j}] = i\beta \, \epsilon_{ij}{}^{k} \hat{\mathcal{S}}'_{k} \, , 
\end{equation}
%2
\begin{equation} \label{rearrange-S.2}
[\hat{\mathcal{S}}_{i},\hat{\mathcal{S}}'_{j}] = [\hat{\mathcal{S}}_{j},\hat{\mathcal{S}}'_{i}] = 0 \, , 
\end{equation}
%3
\begin{equation} \label{rearrange-S.3}
[\hat{\mathcal{S}}_{i},\hat{\mathbb{S}}_{j}] - [\hat{\mathcal{S}}_{j},\hat{\mathbb{S}}_{i}] = i\hbar \, \epsilon_{ij}{}^{k} \hat{\mathbb{S}}_{k} \, , 
\end{equation}
%4
\begin{equation} \label{rearrange-S.4}
[\hat{\mathcal{S}}'_{i},\hat{\mathbb{S}}_{j}] - [\hat{\mathcal{S}}'_{j},\hat{\mathbb{S}}_{i}] = i\beta \, \epsilon_{ij}{}^{k} \hat{\mathbb{S}}_{k} \, , 
\end{equation}
%5
\begin{equation} \label{rearrange-S.5}
[\hat{\mathbb{S}}_{i},\hat{\mathbb{S}}_{j}] = i\beta \, \epsilon_{ij}{}^{k}\hat{\mathcal{S}}_{k} + i\hbar \, \epsilon_{ij}{}^{m}\hat{\mathcal{S}}'_{m} \, ,
\end{equation}
\end{subequations}
and 
\begin{subequations} \label{Clifford-S}
%1
\begin{equation} \label{Clifford-S.1}
\left\{\hat{\mathcal{S}}_{i},\hat{\mathcal{S}}_{j}\right\} = \frac{\hbar^2}{2} \, \delta_{ij} \mathbb{I} \, , \quad \left\{\hat{\mathcal{S}}'_{i},\hat{\mathcal{S}}'_{j}\right\} = \frac{\beta^2}{2} \, \delta_{ij} \mathbb{I} \, , 
\end{equation}
%2
%\begin{equation} \label{Clifford-S.2}
%\left\{\hat{\mathcal{S}}_{i},\hat{\mathcal{S}}'_{j}\right\} + \left\{\hat{\mathcal{S}}_{j},\hat{\mathcal{S}}'_{i}\right\} = 0 \, , 
%\end{equation}
%3
\begin{equation} \label{Clifford-S.3}
\left\{\hat{\mathcal{S}}_{i},\hat{\mathbb{S}}_{j}\right\} + \left\{\hat{\mathcal{S}}_{j},\hat{\mathbb{S}}_{i}\right\} = 0 \, , 
\end{equation}
%4
\begin{equation} \label{Clifford-S.4}
\left\{\hat{\mathcal{S}}'_{i},\hat{\mathbb{S}}_{j}\right\} + \left\{\hat{\mathcal{S}}'_{j},\hat{\mathbb{S}}_{i}\right\} = 0 \, , 
\end{equation}
%5
\begin{equation} \label{Clifford-S.5}
\left\{\hat{\mathbb{S}}_{i},\hat{\mathbb{S}}_{j}\right\} = \hbar\beta \, \delta_{ij} \mathbb{I} - \left\{\hat{\mathcal{S}}_{i},\hat{\mathcal{S}}'_{j}\right\} - \left\{\hat{\mathcal{S}}_{j},\hat{\mathcal{S}}'_{i}\right\} \, ,
\end{equation}
\end{subequations}
where $\mathbb{I}$ denotes the four-dimensional identity matrix and $\left\{ \, . \, , \, . \, \right\}$ is the anticommutator. 
Choosing $s = s' = 1/2$, or, more specifically, $s = \hbar/2$ and $s' = \beta/2$, corresponds to the presence of canonical spin-1/2 fermions (e.g., electrons) in the smeared background \cite{Lake:2019nmn,Lake:2020rwc}. 
(Throughout the rest of this paper, we restrict ourselves to this scenario, which is of greatest physical interest. Higher-spin structures will be analysed, in depth, in a later publication.) 

Equations (\ref{rearrange-S.1})-(\ref{rearrange-S.5}) then combine to give the rescaled Lie algebra
\begin{subequations} \label{}
\begin{eqnarray} \label{SS_commutator}
[\hat{{\rm S}}_{i},\hat{{\rm S}}_{j}] = i(\hbar + \beta) \epsilon_{ij}{}^{k}\hat{\rm S}_{k} \, ,
\end{eqnarray}
\begin{eqnarray} \label{S^2S_commutator}
[\hat{{\rm S}}_{i},\hat{{\rm S}}^{2}] = 0 \, , 
\end{eqnarray}
\end{subequations}
and Eqs. (\ref{Clifford-S.1})-(\ref{Clifford-S.5}) combine to give the rescaled Clifford algebra
\begin{eqnarray} \label{SS_anticommutator}
\left\{\hat{{\rm S}}_{i},\hat{{\rm S}}_{j}\right\} = \frac{(\hbar + \beta)^2}{2} \delta_{ij} \, \mathbb{I} \, .
\end{eqnarray}

Written explicitly, the generalised spin operators take the form
\begin{eqnarray} \label{S_i_explicit}
&&\hat{{\rm S}}_{x} = 
\begin{bmatrix}
    0  &  \frac{(\beta + i\sqrt{\hbar\beta})}{2} & \frac{(\hbar - i\sqrt{\hbar\beta})}{2}  &  0 \\
    \frac{(\beta - i\sqrt{\hbar\beta})}{2}  &  0 & 0  &  \frac{(\hbar + i\sqrt{\hbar\beta})}{2} \\
    \frac{(\hbar + i\sqrt{\hbar\beta})}{2} &  0 & 0  &  \frac{(\beta - i\sqrt{\hbar\beta})}{2} \\
    0  &  \frac{(\hbar - i\sqrt{\hbar\beta})}{2}  & \frac{(\beta + i\sqrt{\hbar\beta})}{2} &  0 
\end{bmatrix}
\, , \nonumber\\ 
&&\hat{{\rm S}}_{y} = 
\begin{bmatrix}
    0  &  -\frac{(i\beta - \sqrt{\hbar\beta})}{2} & -\frac{(i\hbar + \sqrt{\hbar\beta})}{2}  &  0 \\
    \frac{(i\beta + \sqrt{\hbar\beta})}{2}  &  0 & 0  &  -\frac{(i\hbar - \sqrt{\hbar\beta})}{2} \\
    \frac{(i\hbar - \sqrt{\hbar\beta})}{2} &  0 & 0  &  -\frac{(i\beta + \sqrt{\hbar\beta})}{2} \\
    0  &  \frac{(i\hbar + \sqrt{\hbar\beta})}{2}  & \frac{(i\beta - \sqrt{\hbar\beta})}{2}  &  0 
\end{bmatrix}
\, , \nonumber\\
&&\hat{{\rm S}}_{z} =
\begin{bmatrix}
    \frac{(\hbar + \beta)}{2}  &  0  &  0  &  0 \\
    0  &  \frac{(\hbar - \beta)}{2}  &  i\sqrt{\hbar\beta}  &  0 \\
    0  &  -i\sqrt{\hbar\beta}  &  -\frac{(\hbar - \beta)}{2}  &  0 \\
    0  &  0  &  0  &  -\frac{(\hbar+\beta)}{2} 
\end{bmatrix}
\, ,
\end{eqnarray}
and the spin-squared operator is
\begin{eqnarray}  \label{S^2_explicit}
\hat{{\rm S}}^{2} = \sum_{i=1}^{3} \hat{\rm S}_{i}^2 = \frac{3(\hbar + \beta)^2}{4} \, \mathbb{I} \, . 
\end{eqnarray}

The normalized eigenvectors of $\hat{S}_{\rm z}$ are
\begin{subequations} \label{|z_up_down>}
\begin{eqnarray} \label{|up_z>}
\ket{\ket{\uparrow_{\rm z}}} = \ket{\uparrow_{\rm z}}_1 \ket{\uparrow_{\rm z}}_2 = (1,0,0,0) \, ,
\end{eqnarray}
\begin{eqnarray} \label{|down_z>}
\ket{\ket{\downarrow_{\rm z}}} = \ket{\downarrow_{\rm z}}_1 \ket{\downarrow_{\rm z}}_2 = (0,0,0,1) \, ,
\end{eqnarray}
\end{subequations}
and 
\begin{subequations} \label{|z_up_down>'}
\begin{eqnarray} \label{|up_z>'}
\ket{\ket{\uparrow'_{\rm z}}} &=& \frac{1}{\sqrt{1+\delta}}(\ket{\uparrow_{\rm z}}_1 \ket{\downarrow_{\rm z}}_2 -i\sqrt{\delta}\ket{\downarrow_{\rm z}}_1 \ket{\uparrow_{\rm z}}_2)
\nonumber\\
&=& \frac{1}{\sqrt{1+\delta}}(0,1,-i\sqrt{\delta},0) \, ,
\end{eqnarray}
\begin{eqnarray} \label{|down_z>'}
\ket{\ket{\downarrow'_{\rm z}}} &=& \frac{1}{\sqrt{1+\delta}}(\ket{\downarrow_{\rm z}}_1 \ket{\uparrow_{\rm z}}_2 -i\sqrt{\delta}\ket{\uparrow_{\rm z}}_1 \ket{\downarrow_{\rm z}}_2)
\nonumber\\
&=& \frac{1}{\sqrt{1+\delta}}(0,-i\sqrt{\delta},1,0) \, ,
\end{eqnarray}
\end{subequations}
where we have again used the dimensionless parameter $\delta = \beta/\hbar \simeq 10^{-61}$ (\ref{delta}). 
The double ket notation $\ket{\ket{ \dots }}$, introduced in Eqs. (\ref{|up_z>})-(\ref{|down_z>}) and (\ref{|up_z>'})-(\ref{|down_z>'}), is intended to emphasise that spin `up' and spin `down' states in the smeared space theory are, in fact, states in the tensor product Hilbert space $\mathcal{H} = \mathcal{H}_1 \otimes \mathcal{H}_2$. 
Throughout the rest of this work we use double kets to distinguish smeared spin states from their canonical counterparts.

A few comments are in order. 
First, we note that, despite their radically different structures, the eigenvectors $\ket{\ket{\uparrow_{\rm z}}}$ and $\ket{\ket{\downarrow_{\rm z}}}$ are empirically {\it indistinguishable} from the eigenvectors $\ket{\ket{\uparrow'_{\rm z}}}$ and $\ket{\ket{\downarrow'_{\rm z}}}$, respectively, via simultaneous measurements of $\hat{{\rm S}}_{z}$ and $\hat{{\rm S}}^2$. 
For this reason, notation of the form $\ket{\ket{\hat{{\rm S}}_{z} = \dots ,\hat{{\rm S}}^{2} = \dots}}$ is not useful to distinguish between eigenstates with degenerate eigenvalues. 
This also motivates the introduction of new notation for the `canonical' and `geometric' qubit states, namely
\begin{subequations} \label{|4_qubit}
\begin{eqnarray} \label{canonical_qubits}
\ket{\ket{ 0 }} = \ket{\ket{ \uparrow_{\rm z} }} \, , \quad \ket{\ket{ 1 }} = \ket{\ket{ \downarrow_{\rm z} }} \, , 
\end{eqnarray}
\begin{eqnarray} \label{gravitational_qubits}
\ket{\ket{ 0' }} = \ket{\ket{ \uparrow'_{\rm z} }} \, , \quad \ket{\ket{ 1' }} = \ket{\ket{ \downarrow'_{\rm z} }} \, , 
\end{eqnarray}
\end{subequations}
as promised in the Introduction. 
In contrast to canonical quantum mechanics, the measurement of a spin `up' state in smeared space is not sufficient to collapse the wave vector into a single eigenstate of $\hat{{\rm S}}_{\rm z}$. 
Instead, this measurement yields a two-parameter family of states, 
\begin{eqnarray} \label{z-spin_up_superposition}
\ket{\ket{\hat{{\rm S}}_{\rm z} = +(\hbar + \beta)/2,\hat{{\rm S}}^{2} = 3(\hbar + \beta)^2/4}} 
%\nonumber\\
= \alpha_{\rm z} \ket{\ket{\uparrow_{\rm z}}}  + \alpha'_{\rm z} \ket{\ket{\uparrow'_{\rm z}}} \, , 
\end{eqnarray}
where $|\alpha_{\rm z}|^2+|\alpha'_{\rm z}|^2 = 1$. 
Similarly, a spin `down' measurement yields
\begin{eqnarray} \label{z-spin_down_superposition}
\ket{\ket{\hat{{\rm S}}_{\rm z} = -(\hbar + \beta)/2,\hat{{\rm S}}^{2} = 3(\hbar + \beta)^2/4}} 
%\nonumber\\
= \tilde{\alpha}_{\rm z} \ket{\ket{\downarrow_{\rm z}}}  + \tilde{\alpha}'_{\rm z} \ket{\ket{\downarrow'_{\rm z}}} \, , 
\end{eqnarray}
where $|\tilde{\alpha}_{\rm z}|^2+|\tilde{\alpha}'_{\rm z}|^2 = 1$. 
The most general pre-measurement state is given by
\begin{eqnarray} \label{pre-measurement_Psi_spin}
\ket{\ket{\Psi}}_{\rm spin} = \alpha_{\rm z} \ket{\ket{\uparrow_{\rm z}}}  + \alpha'_{\rm z} \ket{\ket{\uparrow'_{\rm z}}} + \tilde{\alpha}_{\rm z} \ket{\ket{\downarrow_{\rm z}}}  + \tilde{\alpha}'_{\rm z} \ket{\ket{\downarrow'_{\rm z}}} \, ,
\end{eqnarray}
where the coefficients are instead subject to the constraint $|\alpha_{\rm z}|^2 + |\alpha'_{\rm z}|^2 + |\tilde{\alpha}_{\rm z}|^2 + |\tilde{\alpha}'_{\rm z}|^2$ = 1.
We refer to the qubits (\ref{canonical_qubits}) as `canonical' because they are simple products of the true canonical qubits, $\ket{\uparrow_{\rm z}}$ and $\ket{\downarrow_{\rm z}}$, and to (\ref{gravitational_qubits}) as `geometric' because they are entangled with the quantum state of the background geometry. 
However, as we will now show, all four eigenstates of both $\hat{{\rm S}}_{\rm y}$ and $\hat{{\rm S}}_{\rm x}$ are entangled states. 
In this sense, there are only two canonical-type qubits in this model.

The normalised eigenstates of $\hat{\rm S}_{\rm y}$ and $\hat{\rm S}_{\rm x}$ may be written as
%\begin{widetext}
\begin{subequations} \label{|y_up_down>}
\begin{eqnarray} \label{|up_y>}
\ket{\ket{\uparrow_{\rm y}}} &=& \frac{1}{\sqrt{2} \sqrt{1+\delta}}(\ket{\uparrow_{\rm y}}_1\ket{\uparrow_{\rm y}}_2 - i\sqrt{\delta}\ket{\downarrow_{\rm y}}_1\ket{\uparrow_{\rm y}}_2 + (1-i\sqrt{\delta})\ket{\uparrow_{\rm y}}_1\ket{\downarrow_{\rm y}}_2) 
\nonumber\\
&=& \frac{1}{\sqrt{2}\sqrt{1+\delta}}(-1+i\sqrt{\delta},-\sqrt{\delta},-i,0) \, ,
\end{eqnarray}
\begin{eqnarray} \label{|down_y>}
\ket{\ket{\downarrow_{\rm y}}} &=& \frac{1}{\sqrt{2} \sqrt{1+\delta}}(\ket{\downarrow_{\rm y}}_1\ket{\downarrow_{\rm y}}_2 - i\sqrt{\delta}\ket{\uparrow_{\rm y}}_1\ket{\downarrow_{\rm y}}_2 + (1-i\sqrt{\delta})\ket{\downarrow_{\rm y}}_1\ket{\uparrow_{\rm y}}_2) 
\nonumber\\
&=& \frac{1}{\sqrt{2}\sqrt{1+\delta}}(1-i\sqrt{\delta},-\sqrt{\delta},-i,0) \, ,
\end{eqnarray}
\end{subequations}
\begin{subequations} \label{|y_up_down>'}
\begin{eqnarray} \label{|up_y>'}
\ket{\ket{\uparrow'_{\rm y}}} &=& \frac{1}{\sqrt{2} \sqrt{1+\delta}}(\ket{\uparrow_{\rm y}}_1\ket{\downarrow_{\rm y}}_2 - i\sqrt{\delta}\ket{\downarrow_{\rm y}}_1\ket{\uparrow_{\rm y}}_2 - (1+i\sqrt{\delta})\ket{\uparrow_{\rm y}}_1\ket{\uparrow_{\rm y}}_2) 
\nonumber\\
&=& \frac{1}{\sqrt{2}\sqrt{1+\delta}}(-i\sqrt{\delta},-i+\sqrt{\delta},0,1) \, ,
\end{eqnarray}
\begin{eqnarray} \label{|down_y>'}
\ket{\ket{\downarrow'_{\rm y}}} &=& \frac{1}{\sqrt{2} \sqrt{1+\delta}}(\ket{\downarrow_{\rm y}}_1\ket{\uparrow_{\rm y}}_2 - i\sqrt{\delta}\ket{\uparrow_{\rm y}}_1\ket{\downarrow_{\rm y}}_2 - (1+i\sqrt{\delta})\ket{\downarrow_{\rm y}}_1\ket{\downarrow_{\rm y}}_2) 
\nonumber\\
&=& \frac{1}{\sqrt{2}\sqrt{1+\delta}}(-i\sqrt{\delta},i-\sqrt{\delta},0,1) \, ,
\end{eqnarray}
\end{subequations}
and 
\begin{subequations} \label{|x_up_down>}
\begin{eqnarray} \label{|up_x>}
\ket{\ket{\uparrow_{\rm x}}} &=& \frac{1}{\sqrt{2} \sqrt{1+\delta}} (\ket{\uparrow_{\rm x}}_1\ket{\downarrow_{\rm x}}_2 - i\sqrt{\delta}\ket{\downarrow_{\rm x}}_1\ket{\uparrow_{\rm x}}_2 + (1-i\sqrt{\delta})\ket{\uparrow_{\rm x}}_1\ket{\uparrow_{\rm x}}_2) 
\nonumber\\
&=& \frac{1}{\sqrt{2} \sqrt{1+\delta}} (1-i\sqrt{\delta},-i\sqrt{\delta},1,0) \, ,
\end{eqnarray}
\begin{eqnarray} \label{|down_x>}
\ket{\ket{\downarrow_{\rm x}}} &=& \frac{1}{\sqrt{2} \sqrt{1+\delta}} (\ket{\downarrow_{\rm x}}_1\ket{\uparrow_{\rm x}}_2 - i\sqrt{\delta}\ket{\uparrow_{\rm x}}_1\ket{\downarrow_{\rm x}}_2 + (1-i\sqrt{\delta})\ket{\downarrow_{\rm x}}_1\ket{\downarrow_{\rm x}}_2) 
\nonumber\\
&=& \frac{1}{\sqrt{2} \sqrt{1+\delta}} (1-i\sqrt{\delta},i\sqrt{\delta},-1,0) \, ,
\end{eqnarray}
\end{subequations}
\begin{subequations} \label{|x_up_down>'}
\begin{eqnarray} \label{|up_x>'}
\ket{\ket{\uparrow'_{\rm x}}} &=& \frac{1}{\sqrt{2} \sqrt{1+\delta}} (\ket{\uparrow_{\rm x}}_1\ket{\downarrow_{\rm x}}_2 - i\sqrt{\delta}\ket{\downarrow_{\rm x}}_1\ket{\uparrow_{\rm x}}_2 - (1+i\sqrt{\delta})\ket{\uparrow_{\rm x}}_1\ket{\uparrow_{\rm x}}_2)  
\nonumber\\
&=& \frac{1}{\sqrt{2} \sqrt{1+\delta}} (-i\sqrt{\delta},-1-i\sqrt{\delta},0,-1) \, ,
\end{eqnarray}
\begin{eqnarray} \label{|down_x>'}
\ket{\ket{\downarrow'_{\rm x}}} &=& \frac{1}{\sqrt{2} \sqrt{1+\delta}} (\ket{\downarrow_{\rm x}}_1\ket{\uparrow_{\rm x}}_2 - i\sqrt{\delta}\ket{\uparrow_{\rm x}}_1\ket{\downarrow_{\rm x}}_2 - (1+i\sqrt{\delta})\ket{\downarrow_{\rm x}}_1\ket{\downarrow_{\rm x}}_2)  
\nonumber\\
&=& \frac{1}{\sqrt{2} \sqrt{1+\delta}} (-i\sqrt{\delta},1+i\sqrt{\delta},0,-1) \, ,
\end{eqnarray}
\end{subequations}
%\end{widetext}
respectively. 
However, most importantly, they can also be rewritten in terms of the $z$-spin eigenstates as
%\begin{widetext}
\begin{subequations} \label{|y_up_down>*}
\begin{eqnarray} \label{|up_y>*}
\ket{\ket{\uparrow_{\rm y}}} = \frac{1}{\sqrt{2}}\left[\frac{-1+i\sqrt{\delta}}{\sqrt{1+\delta}}\ket{\ket{ \uparrow_{\rm z} }} - i\ket{\ket{ \downarrow'_{\rm z} }}\right] \, ,
\end{eqnarray}
\begin{eqnarray} \label{|down_y>*}
\ket{\ket{\downarrow_{\rm y}}} = \frac{1}{\sqrt{2}}\left[-\frac{-1+i\sqrt{\delta}}{\sqrt{1+\delta}}\ket{\ket{ \uparrow_{\rm z} }} - i\ket{\ket{ \downarrow'_{\rm z} }}\right] \, ,
\end{eqnarray}
\end{subequations}
\begin{subequations} \label{|y_up_down>'*}
\begin{eqnarray} \label{|up_y>'*}
\ket{\ket{\uparrow'_{\rm y}}} = \frac{1}{\sqrt{2}\sqrt{1+\delta}}\left[- i\sqrt{\delta}\ket{\ket{ \uparrow_{\rm z} }} + \ket{\ket{ \downarrow_{\rm z} }} - \frac{i-\sqrt{\delta}}{\sqrt{1+\delta}}\left(\ket{\ket{ \uparrow'_{\rm z} }} + i\sqrt{\delta}\ket{\ket{ \downarrow'_{\rm z} }}\right)\right] \, ,
\end{eqnarray}
\begin{eqnarray} \label{|down_y>'*}
\ket{\ket{\downarrow'_{\rm y}}} = \frac{1}{\sqrt{2}\sqrt{1+\delta}}\left[ - i\sqrt{\delta}\ket{\ket{ \uparrow_{\rm z} }} + \ket{\ket{ \downarrow_{\rm z} }} + \frac{i-\sqrt{\delta}}{\sqrt{1+\delta}}\left(\ket{\ket{ \uparrow'_{\rm z} }} + i\sqrt{\delta}\ket{\ket{ \downarrow'_{\rm z} }}\right)\right] \, ,
\end{eqnarray}
\end{subequations}
and 
\begin{subequations} \label{|x_up_down>*}
\begin{eqnarray} \label{|up_x>*}
\ket{\ket{\uparrow_{\rm x}}} = \frac{1}{\sqrt{2}}\left[\frac{1-i\sqrt{\delta}}{\sqrt{1+\delta}}\ket{\ket{ \uparrow_{\rm z} }} + \ket{\ket{ \downarrow'_{\rm z} }}\right] \, ,
\end{eqnarray}
\begin{eqnarray} \label{|down_x>*}
\ket{\ket{\downarrow_{\rm x}}} = \frac{1}{\sqrt{2}}\left[\frac{1-i\sqrt{\delta}}{\sqrt{1+\delta}}\ket{\ket{ \uparrow_{\rm z} }} - \ket{\ket{ \downarrow'_{\rm z} }}\right] \, ,
\end{eqnarray}
\end{subequations}
\begin{subequations} \label{|x_up_down>'*}
\begin{eqnarray} \label{|up_x>'*}
\ket{\ket{\uparrow'_{\rm x}}} = \frac{1}{\sqrt{2}\sqrt{1+\delta}}\left[-i\sqrt{\delta}\ket{\ket{ \uparrow_{\rm z} }} - \ket{\ket{ \downarrow_{\rm z} }} - \frac{1+i\sqrt{\delta}}{\sqrt{1+\delta}}\left(\ket{\ket{ \uparrow'_{\rm z} }} + i\sqrt{\delta}\ket{\ket{ \downarrow'_{\rm z} }} \right)\right] \, ,
\end{eqnarray}
\begin{eqnarray} \label{|down_x>'*}
\ket{\ket{\downarrow'_{\rm x}}} = \frac{1}{\sqrt{2}\sqrt{1+\delta}}\left[-i\sqrt{\delta}\ket{\ket{ \uparrow_{\rm z} }} - \ket{\ket{ \downarrow_{\rm z} }} + \frac{1+i\sqrt{\delta}}{\sqrt{1+\delta}}\left(\ket{\ket{ \uparrow'_{\rm z} }} + i\sqrt{\delta}\ket{\ket{ \downarrow'_{\rm z} }} \right)\right] \, .
\end{eqnarray}
\end{subequations}
%\end{widetext}
Equations (\ref{|y_up_down>*})-(\ref{|y_up_down>'*}) should be compared with the canonical expressions (\ref{y_up_down}), whereas (\ref{|x_up_down>*})-(\ref{|x_up_down>'*}) should be compared with (\ref{x_up_down}). 

We can now consider successive spin measurements along different axes, in the smeared space theory, and compare them with their counterparts in canonical quantum mechanics.  
To this end, we define the general post-measurement states as
\begin{eqnarray} \label{i-spin_up_superposition}
\ket{\ket{\hat{{\rm S}}_{i} = +(\hbar + \beta)/2,\hat{{\rm S}}^{2} = 3(\hbar + \beta)^2/4}}  
%\nonumber\\
= \alpha_{i} \ket{\ket{\uparrow_{i}}}  + \alpha'_{i} \ket{\ket{\uparrow'_{i}}} \, , 
\end{eqnarray}
where $|\alpha_{i}|^2+|\alpha'_{i}|^2 = 1$, and 
\begin{eqnarray} \label{i-spin_down_superposition}
\ket{\ket{\hat{{\rm S}}_{i} = -(\hbar + \beta)/2,\hat{{\rm S}}^{2} = 3(\hbar + \beta)^2/4}}  
%\nonumber\\
= \tilde{\alpha}_{i} \ket{\ket{\downarrow_{i}}}  + \tilde{\alpha}'_{i} \ket{\ket{\downarrow'_{i}}} \, , 
\end{eqnarray}
where $|\tilde{\alpha}_{i}|^2+|\tilde{\alpha}'_{i}|^2 = 1$. 
We then define $P(\uparrow_{i}|\uparrow_{j})$ as the probability that, if the first measurement along the $i$-axis yields spin `up', then the second measurement along the $j$-axis will also yield spin `up'. 
The probabilities $P(\uparrow_{i}|\downarrow_{j})$, $P(\downarrow_{i}|\uparrow_{j})$ and $P(\downarrow_{i}|\downarrow_{j})$ are defined in like manner. 
Using Eqs. (\ref{|y_up_down>*})-(\ref{|y_up_down>'*}) and (\ref{|x_up_down>*})-(\ref{|x_up_down>'*}), it is straightforward to verify that 
\begin{eqnarray} \label{}
P(\uparrow_{i}|\uparrow_{j}) &=& P(\uparrow_{i}|\downarrow_{j}) = 1/2 \, ,
\nonumber\\
P(\downarrow_{i}|\uparrow_{j}) &=& P(\downarrow_{i}|\downarrow_{j}) = 1/2 \, ,
\end{eqnarray}
for all $i \neq j$. 

Despite the very different mathematical structures of the generalised spin operators, their eigenstates remain indistinguishable from the canonical spin eigenstates, via simultaneous measurements of $\hat{{\rm S}}_{i}$ and $\hat{{\rm S}}^2$, except for a very small rescaling of the measured spin values such that $\pm \hbar/2 \rightarrow \pm(\hbar + \beta)/2$. 
Nonetheless, the model is in principle distinguishable from canonical quantum mechanics, even in the spin sector, via the presence of spin GURs. 
These are analogous to the GURs for smeared orbital angular momentum measurements, i.e., 
\begin{eqnarray} \label{spin_GURs}
\Delta_{\Psi}{\rm S}_{i} \, \Delta_{\Psi}{\rm S}_{j} \geq \dots \geq \left(\frac{\hbar + \beta}{2}\right) |\epsilon_{ij}{}^{k} \braket{\hat{{\rm S}}_{k}}_{\Psi}| \, ,
\end{eqnarray}
where the sum of terms in the middle of the two inequalities is determined by the relation
\begin{eqnarray} \label{DS^2*}
(\Delta_{\Psi}{\rm S}_{i})^2 &=& (\Delta_{\Psi}\mathcal{S}_{i})^2 + (\Delta_{\Psi}\mathcal{S}'_{i})^2 + (\Delta_{\Psi}\mathbb{S}_{i})^2 
\nonumber\\
&+& {\rm cov}(\hat{\mathcal{S}}_{i},\hat{\mathbb{S}}_{i}) + {\rm cov}(\hat{\mathbb{S}}_{i},\hat{\mathcal{S}}_{i})
\nonumber\\
&+& {\rm cov}(\hat{\mathcal{S}}'_{i},\hat{\mathbb{S}}_{i}) + {\rm cov}(\hat{\mathbb{S}}_{i},\hat{\mathcal{S}}'_{i}) \, .
\end{eqnarray}

We now introduce the smeared space creation and annihilation operators,
\begin{eqnarray} \label{Sigma_pm}
\hat{{\rm S}}_{\pm} = \hat{{\rm S}}_{\rm x} \pm i\hat{{\rm S}}_{\rm y} \, , 
\end{eqnarray}
which satisfy the algebra
%\begin{subequations} \label{S_pm_Lie_algebra}
\begin{eqnarray} \label{S_pm_Lie_algebra}
%
%\begin{eqnarray} \label{S_pm_Lie_algebra-A}
[\hat{{\rm S}}_{\rm z},\hat{{\rm S}}_{\pm}] = \pm (\hbar + \beta) \, \hat{{\rm S}}_{\pm} \, , \quad
%\end{eqnarray}
%
%\begin{eqnarray} \label{S_pm_Lie_algebra-B}
[\hat{{\rm S}}_{+},\hat{{\rm S}}_{-}] = 2(\hbar + \beta) \, \hat{{\rm S}}_{\rm z} \, .
%\end{eqnarray}
\end{eqnarray}
%\end{subequations} 
Written explicitly, $\hat{{\rm S}}_{\pm}$ take the form
\begin{eqnarray} \label{S_pm_explicit}
\hat{{\rm S}}_{+} &=& 
\begin{bmatrix}
    0 & \beta + i\sqrt{\hbar\beta} & \hbar - i\sqrt{\hbar\beta} & 0 \\
    0 & 0 & 0 & \hbar + i\sqrt{\hbar\beta} \\
    0 & 0 & 0 & \beta - i\sqrt{\hbar\beta} \\
    0 & 0 & 0 & 0  
\end{bmatrix}
\, , \nonumber\\ 
\hat{{\rm S}}_{-} &=& 
\begin{bmatrix}
    0 & 0 & 0 & 0 \\
    \beta - i\sqrt{\hbar\beta} & 0 & 0 & 0 \\
    \hbar + i\sqrt{\hbar\beta} & 0 & 0 & 0 \\
    0 & \hbar - i\sqrt{\hbar\beta} & \beta + i\sqrt{\hbar\beta} & 0  
\end{bmatrix}
\, .
\end{eqnarray}

The eigenvectors of $\hat{{\rm S}}_{+}$ are two copies of the null vector, $\ket{\ket{\rm null}} = (0,0,0,0)$, plus the $z$-spin up eigenstates $\ket{\ket{\uparrow_{\rm z}}} = (1,0,0,0)$ (\ref{|up_z>}) and $\ket{\ket{\uparrow'_{\rm z}}} = (1+\delta)^{-1/2}(0,1,-i\sqrt{\delta},0)$ (\ref{|up_z>'}), whereas the eigenvectors of $\hat{{\rm S}}_{-}$ are two copies of the null vector, plus the $z$-spin down eigenstates $\ket{\ket{\downarrow_{\rm z}}} = (0,0,0,1)$ (\ref{|down_z>}) and $\ket{\ket{\downarrow'_{\rm z}}} = (1+\delta)^{-1/2}(0,-i\sqrt{\delta},1,0)$ (\ref{|down_z>'}). 
For each operator, all four eigenvectors correspond to the eigenvalue $0$. 
Hence, $\hat{{\rm S}}_{\pm}$ are analogous to their canonical counterparts (\ref{s_pm_explicit}), but each annihilates two of the four spin eigenstates of $\hat{{\rm S}}_{\rm z}$.   

The generalised creation and annihilation operators perform spin flips according to:
\begin{subequations} \label{spin_flip_z_z'}
\begin{eqnarray} \label{spin_flip_z}
\hat{{\rm S}}_{-} \ket{\ket{\uparrow_{\rm z}}} &=& \sqrt{1+\delta} \, (\hbar + i\sqrt{\hbar\beta})\ket{\ket{\downarrow'_{\rm z}}} \, , 
\nonumber\\ 
\hat{{\rm S}}_{-} \ket{\ket{\uparrow'_{\rm z}}} &=& \sqrt{1+\delta} \, (\hbar - i\sqrt{\hbar\beta}) \ket{\ket{\downarrow_{\rm z}}} \, ,
\end{eqnarray}
\begin{eqnarray} \label{spin_flip_z'}
\hat{{\rm S}}_{+} \ket{\ket{\downarrow_{\rm z}}} &=& \sqrt{1+\delta} \, (\hbar + i\sqrt{\hbar\beta}) \ket{\ket{\uparrow'_{\rm z}}} \, , 
\nonumber\\ 
\hat{{\rm S}}_{+} \ket{\ket{\downarrow'_{\rm z}}} &=& \sqrt{1+\delta} \, (\hbar - i\sqrt{\hbar\beta}) \ket{\ket{\uparrow_{\rm z}}} \, .
\end{eqnarray}
\end{subequations} 
These may be compared with Eqs. (\ref{spin_flip_z**}) in the canonical theory. 
It is interesting to note that $\hat{\rm S}_{\pm}$ flip unentangled states to entangled states with opposite spin, and vice versa, and that the corresponding dimensionful values are complex conjugates of one another.

Finally, before concluding this section, we note that the generalised smeared space qubits $\left\{\ket{\ket{\uparrow_{\rm z}}},\ket{\ket{\downarrow_{\rm z}}};\ket{\ket{\uparrow'_{\rm z}}},\ket{\ket{\downarrow'_{\rm z}}}\right\}$ satisfy the braket relations
\begin{subequations} \label{generalised_braket_relations}
\begin{eqnarray} \label{generalised_braket_relations-1}
\braket{\braket{ \uparrow_{i} || \uparrow_{i} }} &=& 1 \, , \quad \braket{\braket{ \uparrow_{i} || \downarrow_{i} }} = 0 \, , 
\nonumber\\
\braket{\braket{ \downarrow_{i} || \uparrow_{i} }} &=& 0 \, , \quad \braket{\braket{ \downarrow_{i} || \downarrow_{i} }} = 1 \, , 
\end{eqnarray}
\begin{eqnarray} \label{generalised_braket_relations-1}
\braket{\braket{ \uparrow'_{i} || \uparrow'_{i} }} &=& 1 \, , \quad \braket{\braket{ \uparrow'_{i} || \downarrow'_{i} }} = 0 \, , 
\nonumber\\
\braket{\braket{ \downarrow'_{i} || \uparrow'_{i} }} &=& 0 \, , \quad \braket{\braket{ \downarrow'_{i} || \downarrow'_{i} }} = 1 \, , 
\end{eqnarray}
\begin{eqnarray} \label{generalised_braket_relations-1}
\braket{\braket{ \uparrow_{i} || \uparrow'_{i} }} &=& 0 \, , \quad \braket{\braket{ \uparrow_{i} || \downarrow'_{i} }} = 0 \, , 
\nonumber\\
\braket{\braket{ \downarrow_{i} || \uparrow'_{i} }} &=& 0 \, , \quad \braket{\braket{ \downarrow_{i} || \downarrow'_{i} }} = 0 \, . 
\end{eqnarray}
\end{subequations}
These may be compared with Eqs. (\ref{canonical_spin_braket_relations}). 
A key point is that the primed and unprimed eigenvectors are orthogonal in the tensor product Hilbert space, $\mathcal{H}_{1} \otimes \mathcal{H}_{2}$, even when both represent either spin `up' or spin `down' states, according to their measurable eigenvalues. 
This permits expansions of the form (\ref{i-spin_up_superposition}) and (\ref{i-spin_down_superposition}).

%%%%%%%%%%%%%%%%%%%%%%%%%%%%%%%%%%%%%%%%%%%%%%%%%%%%%%%%%%
\subsection{Two-particle systems in smeared space} \label{Sec.3.2}

In order to construct the generalised spin operators for two-particle states, we must carefully consider the physical meaning of the second spin subspace in the tensor product space for single particles, i.e., $\mathcal{H} = \mathcal{H}_1 \otimes \mathcal{H}_2$, where $\ket{\psi}_{\rm spin} \in \mathcal{H}_1$ is the canonical one-particle state and $\ket{g}_{\rm spin} \in \mathcal{H}_2$ determines the influence, on it, of the quantum background geometry. 
If $\ket{g}_{\rm spin}$ is interpreted literally, that is, as the physical spin state of the background, then the correct generalisation to two particles, labelled $A$ and $B$, takes the form $\mathcal{H}_1 \otimes \mathcal{H}_2 \rightarrow \mathcal{H}_{\rm A1} \otimes \mathcal{H}_{\rm B1} \otimes \mathcal{H}_2$. 

However, this is at odds with the treatment of the non-spin part of the multiparticle wave function presented in \cite{Lake:2018zeg}. 
In this work, $N$-particle states were constructed by taking the canonical variables $\left\{\bold{q}_{\rm A},\bold{q}_{\rm B}, \dots \bold{q}_{\rm N}\right\} = \left\{\bold{x}_{\rm A},\bold{x}_{\rm B}, \dots \bold{x}_{\rm N}\right\}$ and adding to them the `geometric' degrees of freedom $\left\{\bold{q}'_{\rm A},\bold{q}'_{\rm B}, \dots \bold{q}'_{\rm N}\right\} = \left\{\bold{x}'_{\rm A}-\bold{x}_{\rm A},\bold{x}'_{\rm B}-\bold{x}_{\rm B}, \dots \bold{x}'_{\rm N}-\bold{x}_{\rm N}\right\}$, by analogy with the one-particle case, Eq. (\ref{new_variables-1}).  
In this interpretation, $\ket{g} \in \mathcal{H}_{\rm A2} \otimes \mathcal{H}_{\rm B2} \dots \otimes \mathcal{H}_{\rm N2}$ does not represent the quantum state of the background per se, but, instead, the ability of the nonlocal geometry to influence any number of particles that may be propagating within it. 
It is therefore reasonable that the form of $\ket{g}$, and the number of subspaces in the Hilbert space to which it belongs, depend on the number of degrees of freedom in the matter sector, i.e., on the number of particles. 
Extending this interpretation to the spin sector, the spin Hilbert space is expanded such that $\mathcal{H}_1 \otimes \mathcal{H}_2 \rightarrow (\mathcal{H}_{\rm A1} \otimes \mathcal{H}_{\rm A2}) \otimes (\mathcal{H}_{\rm B1} \otimes \mathcal{H}_{\rm B2}) \dots \otimes (\mathcal{H}_{\rm N1} \otimes \mathcal{H}_{\rm N2})$ for an $N$-particle state. 

The two-particle spin operators for the smeared space theory are then constructed as
\begin{eqnarray} \label{s^i_AB}
\hat{{\rm S}}^{\rm (A)}_{i} = (\mathcal{{S}}^{\rm (A)}_{i} + \mathcal{{S}}'^{\rm (A)}_{i} + \mathbb{{S}}^{\rm (A)}_{i}) \otimes \mathbb{I}_{\rm B} \, ,
\quad
\hat{{\rm S}}^{\rm (B)}_{i} = \mathbb{I}_{\rm A} \otimes (\mathcal{{S}}^{\rm (B)}_{i} + \mathcal{{S}}'^{\rm (B)}_{i} + \mathbb{{S}}^{\rm (B)}_{i}) \, ,
\end{eqnarray}
where $\mathbb{I}_{\rm A} = (\mathbb{I}_{\rm A1} \otimes \mathbb{I}_{\rm A2})$ and $\mathbb{I}_{\rm B} = (\mathbb{I}_{\rm B1} \otimes \mathbb{I}_{\rm B2})$. 
In this case, each set of operators $\left\{\hat{{\rm S}}^{\rm (A)}_{i}\right\}_{i=1}^{3}$ and $\left\{\hat{{\rm S}}^{\rm (B)}_{i}\right\}_{i=1}^{3}$ satisfies the rescaled Lie algebra (\ref{SS_commutator})-(\ref{S^2S_commutator}) and the two sets commute,
\begin{eqnarray} \label{[S^i_A,S^j_B]}
[\hat{{\rm S}}^{\rm (A)}_{i},\hat{{\rm S}}^{\rm (B)}_{j}] = 0 \, , 
\end{eqnarray}
as in canonical quantum mechanics. 

The total spin in the $i^{\rm th}$ direction is given by
\begin{eqnarray} \label{S_i^(AB)}
\hat{{\rm S}}_{i} = \hat{{\rm S}}^{\rm (A)}_{i} + \hat{{\rm S}}^{\rm (B)}_{i} \, 
\end{eqnarray}
and each $\hat{{\rm S}}_{i}$ has 16 independent eigenvectors. 
These can be written as 4 groups of 4, namely
%\begin{widetext}
\begin{subequations} \label{Two-particle_eigenvectors}
\begin{eqnarray} \label{Two-particle_eigenvectors-1}
\left\{\ket{\ket{\uparrow_{i}}}_{\rm A}\ket{\ket{\uparrow_{i}}}_{\rm B},\ket{\ket{\uparrow_{i}}}_{\rm A}\ket{\ket{\downarrow_{i}}}_{\rm B},\ket{\ket{\downarrow_{i}}}_{\rm A}\ket{\ket{\uparrow_{i}}}_{\rm B},\ket{\ket{\downarrow_{i}}}_{\rm A}\ket{\ket{\downarrow_{i}}}_{\rm B}\right\} \, , 
\end{eqnarray}
\begin{eqnarray} \label{Two-particle_eigenvectors-2}
\left\{\ket{\ket{\uparrow'_{i}}}_{\rm A}\ket{\ket{\uparrow_{i}}}_{\rm B},\ket{\ket{\uparrow'_{i}}}_{\rm A}\ket{\ket{\downarrow_{i}}}_{\rm B},\ket{\ket{\downarrow'_{i}}}_{\rm A}\ket{\ket{\uparrow_{i}}}_{\rm B},\ket{\ket{\downarrow'_{i}}}_{\rm A}\ket{\ket{\downarrow_{i}}}_{\rm B}\right\} \, , 
\end{eqnarray}
\begin{eqnarray} \label{Two-particle_eigenvectors-3}
\left\{\ket{\ket{\uparrow_{i}}}_{\rm A}\ket{\ket{\uparrow'_{i}}}_{\rm B},\ket{\ket{\uparrow_{i}}}_{\rm A}\ket{\ket{\downarrow'_{i}}}_{\rm B},\ket{\ket{\downarrow_{i}}}_{\rm A}\ket{\ket{\uparrow'_{i}}}_{\rm B},\ket{\ket{\downarrow_{i}}}_{\rm A}\ket{\ket{\downarrow'_{i}}}_{\rm B}\right\} \, , 
\end{eqnarray}
\begin{eqnarray} \label{Two-particle_eigenvectors-4}
\left\{\ket{\ket{\uparrow'_{i}}}_{\rm A}\ket{\ket{\uparrow'_{i}}}_{\rm B},\ket{\ket{\uparrow'_{i}}}_{\rm A}\ket{\ket{\downarrow'_{i}}}_{\rm B},\ket{\ket{\downarrow'_{i}}}_{\rm A}\ket{\ket{\uparrow'_{i}}}_{\rm B},\ket{\ket{\downarrow'_{i}}}_{\rm A}\ket{\ket{\downarrow'_{i}}}_{\rm B}\right\} \, . 
\end{eqnarray}
\end{subequations}
%\end{widetext}
The eigenvectors in each group correspond to the eigenvalues $\left\{+(\hbar+\beta),0,0,-(\hbar+\beta)\right\}$, respectively, and may compared with the canonical two-particle eigenstates, Eq. (\ref{s_z_eigenvectors}), which correspond to the eigenvalues $\left\{+\hbar,0,0,-\hbar\right\}$. 
The total spin-squared operator is
\begin{eqnarray} \label{S^2_AB}
\hat{{\rm S}}^{2} = \sum_{i=1}^{3} (\hat{{\rm S}}^{\rm (A)}_{i} + \hat{{\rm S}}^{\rm (B)}_{i})^2 
=  (\hat{\boldsymbol{{\rm S}}}_{\rm A})^2 + 2 \, \hat{\boldsymbol{{\rm S}}}_{\rm A} \, . \, \hat{\boldsymbol{{\rm S}}}_{\rm B} + (\hat{\boldsymbol{{\rm S}}}_{\rm B})^2 \, ,
\end{eqnarray}
where 
\begin{eqnarray} \label{S_AB_vec}
\hat{\bold{{\rm S}}}_{\rm A / B} = \sum_{i=1}^{3} \hat{{\rm S}}^{\rm (A / B)}_{i} \, \bold{e}_{i}(0) \, , 
\end{eqnarray}
and $\left\{\bold{e}_{i}(0)\right\}_{i=1}^{3}$ are the tangent vectors at the coordinate origin. 

From here on, we focus on the simultaneous eigenvectors of the total $z$-spin operator and the total spin-squared operator, as in the standard analysis of two-particle states in canonical quantum mechanics. 
The explicit forms of $\hat{{\rm S}}_{\rm z}$ and $\hat{{\rm S}}^2$ for the smeared two-particle state are given in Appendix \ref{Appendix-B}. 
%Note that, in contrast to the canonical spin-squared operator (\ref{s^2}), $\hat{S}^2$ is asymmetric. 
The 16 independent eigenvectors of $\hat{{\rm S}}_{\rm z}$ and $\hat{{\rm S}}^2$, given in Eqs. (\ref{S_z_AB_explicit}) and (\ref{S^2_AB_explicit}), respectively, can be written in 4 groups of 4, i.e., 
%%
%\begin{widetext}
\begin{subequations} \label{S_i^(AB)_eigenvectors-1}
\begin{eqnarray} \label{S_i^(AB)_eigenvector-1a}
\ket{\ket{\Psi_{\rm 1a}}} &=& \ket{\ket{\uparrow_{\rm z}}}_{\rm A}\ket{\ket{\uparrow_{\rm z}}}_{\rm B} 
\nonumber\\
&=& (1,0,0,0,0,0,0,0,0,0,0,0,0,0,0,0) \, , 
\end{eqnarray} 
\begin{eqnarray} \label{S_i^(AB)_eigenvector-1b}
\ket{\ket{\Psi_{\rm 1b}}} &=& \ket{\ket{\uparrow'_{\rm z}}}_{\rm A}\ket{\ket{\uparrow_{\rm z}}}_{\rm B} 
\nonumber\\
&=& \frac{1}{\sqrt{1+\delta}} (0,0,0,0,1,0,0,0,-i\sqrt{\delta},0,0,0,0,0,0,0) \, , 
\end{eqnarray} 
\begin{eqnarray} \label{S_i^(AB)_eigenvector-1c}
\ket{\ket{\Psi_{\rm 1c}}} &=& \ket{\ket{\uparrow_{\rm z}}}_{\rm A}\ket{\ket{\uparrow'_{\rm z}}}_{\rm B} 
\nonumber\\
&=& \frac{1}{\sqrt{1+\delta}} (0,1,-i\sqrt{\delta},0,0,0,0,0,0,0,0,0,0,0,0,0) \, ,  
\end{eqnarray} 
\begin{eqnarray} \label{S_i^(AB)_eigenvector-1d}
\ket{\ket{\Psi_{\rm 1d}}} &=& \ket{\ket{\uparrow'_{\rm z}}}_{\rm A}\ket{\ket{\uparrow'_{\rm z}}}_{\rm B} 
\nonumber\\
&=& \frac{1}{1+\delta} (0,0,0,0,0,1,-i\sqrt{\delta},0,0,-i\sqrt{\delta},-\delta,0,0,0,0,0) \, , 
\end{eqnarray} 
\end{subequations}
which correspond to the eigenvalues $\hat{{\rm S}}_{\rm z} = +(\hbar+\beta)$, $\hat{{\rm S}}^2 = 2(\hbar+\beta)^2$, 
\begin{subequations} \label{S_i^(AB)_eigenvectors-2}
\begin{eqnarray} \label{S_i^(AB)_eigenvector-2a}
\ket{\ket{\Psi_{\rm 2a}}} &=& \ket{\ket{\downarrow_{\rm z}}}_{\rm A}\ket{\ket{\downarrow_{\rm z}}}_{\rm B} 
\nonumber\\
&=& (0,0,0,0,0,0,0,0,0,0,0,0,0,0,0,1) \, , 
\end{eqnarray} 
\begin{eqnarray} \label{S_i^(AB)_eigenvector-2b}
\ket{\ket{\Psi_{\rm 2b}}} &=& \ket{\ket{\downarrow'_{\rm z}}}_{\rm A}\ket{\ket{\downarrow_{\rm z}}}_{\rm B} 
\nonumber\\
&=& \frac{1}{\sqrt{1+\delta}} (0,0,0,0,0,0,0,-i\sqrt{\delta},0,0,0,1,0,0,0,0) \, , 
\end{eqnarray} 
\begin{eqnarray} \label{S_i^(AB)_eigenvector-2c}
\ket{\ket{\Psi_{\rm 2c}}} &=& \ket{\ket{\downarrow_{\rm z}}}_{\rm A}\ket{\ket{\downarrow'_{\rm z}}}_{\rm B} 
\nonumber\\
&=& \frac{1}{\sqrt{1+\delta}} (0,0,0,0,0,0,0,0,0,0,0,0,0,-i\sqrt{\delta},1,0) \, ,  
\end{eqnarray} 
\begin{eqnarray} \label{S_i^(AB)_eigenvector-2d}
\ket{\ket{\Psi_{\rm 2d}}} &=& \ket{\ket{\downarrow'_{\rm z}}}_{\rm A}\ket{\ket{\downarrow'_{\rm z}}}_{\rm B} 
\nonumber\\
&=& \frac{1}{1+\delta} (0,0,0,0,0,-\delta,-i\sqrt{\delta},0,0,-i\sqrt{\delta},1,0,0,0,0,0)\, , 
\end{eqnarray} 
\end{subequations}
which correspond to $\hat{{\rm S}}_{\rm z} = -(\hbar+\beta)$, $\hat{{\rm S}}^2 = 2(\hbar+\beta)^2$,
\begin{subequations} \label{S_i^(AB)_eigenvectors-3}
\begin{eqnarray} \label{S_i^(AB)_eigenvector-3a}
\ket{\ket{\Psi_{\rm 3a}}} &=& \frac{1}{\sqrt{2}} (\ket{\ket{\uparrow'_{\rm z}}}_{\rm A}\ket{\ket{\downarrow_{\rm z}}}_{\rm B} + \ket{\ket{\downarrow_{\rm z}}}_{\rm A}\ket{\ket{\uparrow'_{\rm z}}}_{\rm B}) 
\nonumber\\
&=& \frac{1}{\sqrt{2}\sqrt{1+\delta}} (0,0,0,0,0,0,0,1,0,0,0,-i\sqrt{\delta},0,1,-i\sqrt{\delta},0) \, , 
\end{eqnarray} 
\begin{eqnarray} \label{S_i^(AB)_eigenvector-3b}
\ket{\ket{\Psi_{\rm 3b}}} &=& \frac{1}{\sqrt{2}} (\ket{\ket{\uparrow_{\rm z}}}_{\rm A}\ket{\ket{\downarrow'_{\rm z}}}_{\rm B} + \ket{\ket{\downarrow'_{\rm z}}}_{\rm A}\ket{\ket{\uparrow_{\rm z}}}_{\rm B}) 
\nonumber\\
&=& \frac{1}{\sqrt{2}\sqrt{1+\delta}} (0,-i\sqrt{\delta},1,0,-i\sqrt{\delta},0,0,0,1,0,0,0,0,0,0,0) \, , 
\end{eqnarray} 
\begin{eqnarray} \label{S_i^(AB)_eigenvector-3c}
\ket{\ket{\Psi_{\rm 3c}}} &=& \frac{1}{\sqrt{2}} \left[ \ket{\ket{\uparrow'_{\rm z}}}_{\rm A}\ket{\ket{\downarrow'_{\rm z}}}_{\rm B} + \frac{(1-i\sqrt{\delta})^2}{1+\delta}\ket{\ket{\downarrow_{\rm z}}}_{\rm A}\ket{\ket{\uparrow_{\rm z}}}_{\rm B}\right] 
\nonumber\\ 
&=& \frac{1}{\sqrt{2}(1+\delta)} (0,0,0,0,0,-i\sqrt{\delta},1,0,0,-\delta,-i\sqrt{\delta},0,(1-i\sqrt{\delta})^2,0,0,0) \, , 
%\ket{\ket{\Psi_{\rm 3c}}} &=& \frac{1}{2} \left[\ket{\ket{\uparrow_{\rm z}}}_{\rm A}\ket{\ket{\downarrow_{\rm z}}}_{\rm B} + \ket{\ket{\downarrow_{\rm z}}}_{\rm A}\ket{\ket{\uparrow_{\rm z}}}_{\rm B} 
%- \frac{(1+i\sqrt{\delta})^2}{1+\delta}\left(\ket{\ket{\uparrow'_{\rm z}}}_{\rm A}\ket{\ket{\downarrow'_{\rm z}}}_{\rm B} + \ket{\ket{\downarrow'_{\rm z}}}_{\rm A}\ket{\ket{\uparrow'_{\rm z}}}_{\rm B}\right)\right] 
%\nonumber\\
%&=& \frac{1}{2(1+\delta)} (0,0,0,1+\delta,0,0,-(1+i\sqrt{\delta})^2,0,0,(1+i\sqrt{\delta})^2,0,0,-(1+\delta),0,0,0) \, , 
\end{eqnarray} 
\begin{eqnarray} \label{S_i^(AB)_eigenvector-3d}
\ket{\ket{\Psi_{\rm 3d}}} &=& \frac{1}{\sqrt{2}} \left[ \ket{\ket{\downarrow'_{\rm z}}}_{\rm A}\ket{\ket{\uparrow'_{\rm z}}}_{\rm B} + \frac{(1-i\sqrt{\delta})^2}{1+\delta}\ket{\ket{\uparrow_{\rm z}}}_{\rm A}\ket{\ket{\downarrow_{\rm z}}}_{\rm B}\right] 
\nonumber\\ 
&=& \frac{1}{\sqrt{2}(1+\delta)} (0,0,0,(1-i\sqrt{\delta})^2,0,-i\sqrt{\delta},-\delta,0,0,1,-i\sqrt{\delta},0,0,0,0,0) \, , 
\end{eqnarray} 
\end{subequations}
which correspond to $\hat{{\rm S}}_{\rm z} = 0$, $\hat{{\rm S}}^2 = 2(\hbar+\beta)^2$, and
\begin{subequations} \label{S_i^(AB)_eigenvectors-4}
\begin{eqnarray} \label{S_i^(AB)_eigenvector-4a}
\ket{\ket{\Phi_{\rm a}}} &=& \frac{1}{\sqrt{2}} (\ket{\ket{\uparrow'_{\rm z}}}_{\rm A}\ket{\ket{\downarrow_{\rm z}}}_{\rm B} - \ket{\ket{\downarrow_{\rm z}}}_{\rm A}\ket{\ket{\uparrow'_{\rm z}}}_{\rm B}) 
\nonumber\\
&=& \frac{1}{\sqrt{2}\sqrt{1+\delta}} (0,0,0,0,0,0,0,1,0,0,0,-i\sqrt{\delta},0,-1,i\sqrt{\delta},0) \, , 
\end{eqnarray} 
\begin{eqnarray} \label{S_i^(AB)_eigenvector-4b}
\ket{\ket{\Phi_{\rm b}}} &=& \frac{1}{\sqrt{2}} (\ket{\ket{\uparrow_{\rm z}}}_{\rm A}\ket{\ket{\downarrow'_{\rm z}}}_{\rm B} - \ket{\ket{\downarrow'_{\rm z}}}_{\rm A}\ket{\ket{\uparrow_{\rm z}}}_{\rm B}) 
\nonumber\\
&=& \frac{1}{\sqrt{2}\sqrt{1+\delta}} (0,-i\sqrt{\delta},1,0,i\sqrt{\delta},0,0,0,-1,0,0,0,0,0,0,0) \, , 
\end{eqnarray} 
\begin{eqnarray} \label{S_i^(AB)_eigenvector-4c}
\ket{\ket{\Phi_{\rm c}}} &=& \frac{1}{\sqrt{2}} \left[ \ket{\ket{\uparrow'_{\rm z}}}_{\rm A}\ket{\ket{\downarrow'_{\rm z}}}_{\rm B} - \frac{(1-i\sqrt{\delta})^2}{1+\delta}\ket{\ket{\downarrow_{\rm z}}}_{\rm A}\ket{\ket{\uparrow_{\rm z}}}_{\rm B}\right] 
\nonumber\\ 
&=& \frac{1}{\sqrt{2}(1+\delta)} (0,0,0,0,0,-i\sqrt{\delta},1,0,0,-\delta,-i\sqrt{\delta},0,-(1-i\sqrt{\delta})^2,0,0,0) \, , 
%\ket{\ket{\Phi_{\rm c}}} &=& \frac{1}{2} \left[\ket{\ket{\uparrow_{\rm z}}}_{\rm A}\ket{\ket{\downarrow_{\rm z}}}_{\rm B} + \ket{\ket{\downarrow_{\rm z}}}_{\rm A}\ket{\ket{\uparrow_{\rm z}}}_{\rm B} 
%+ \frac{(1+i\sqrt{\delta})^2}{1+\delta}\left(\ket{\ket{\uparrow'_{\rm z}}}_{\rm A}\ket{\ket{\downarrow'_{\rm z}}}_{\rm B} + \ket{\ket{\downarrow'_{\rm z}}}_{\rm A}\ket{\ket{\uparrow'_{\rm z}}}_{\rm B}\right)\right] 
%\nonumber\\
%&=& \frac{1}{2(1+\delta)} (0,0,0,1+\delta,0,0,(1+i\sqrt{\delta})^2,0,0,-(1+i\sqrt{\delta})^2,0,0,-(1+\delta),0,0,0) \, , 
\end{eqnarray} 
\begin{eqnarray} \label{S_i^(AB)_eigenvector-4d}
\ket{\ket{\Phi_{\rm d}}} &=& \frac{1}{\sqrt{2}} \left[ \ket{\ket{\downarrow'_{\rm z}}}_{\rm A}\ket{\ket{\uparrow'_{\rm z}}}_{\rm B} - \frac{(1-i\sqrt{\delta})^2}{1+\delta}\ket{\ket{\uparrow_{\rm z}}}_{\rm A}\ket{\ket{\downarrow_{\rm z}}}_{\rm B}\right] 
\nonumber\\
&=& \frac{1}{\sqrt{2}(1+\delta)} (0,0,0,-(1-i\sqrt{\delta})^2,0,-i\sqrt{\delta},-\delta,0,0,1,-i\sqrt{\delta},0,0,0,0,0) \, .
\end{eqnarray} 
\end{subequations}
which correspond to $\hat{{\rm S}}_{\rm z} = 0$, $\hat{{\rm S}}^2 = 0$. 
%\end{widetext}
These may be compared with the canonical two-particle eigenstates (\ref{s_i^(AB)_eigenvector-1})-(\ref{s_i^(AB)_eigenvector-4}). 
The states $\ket{\ket{\Psi_{\rm 1a}}}$ to $\ket{\ket{\Psi_{\rm 1d}}}$, $\ket{\ket{\Psi_{\rm 2a}}}$ to $\ket{\ket{\Psi_{\rm 2d}}}$, $\ket{\ket{\Psi_{\rm 3a}}}$ to $\ket{\ket{\Psi_{\rm 3b}}}$ and $\ket{\ket{\Phi_{\rm a}}}$ to $\ket{\ket{\Phi_{\rm b}}}$ have clear analogues in the canonical theory, but $\ket{\ket{\Psi_{\rm 3c}}}$, $\ket{\ket{\Psi_{\rm 3d}}}$ and $\ket{\ket{\Phi_{\rm c}}}$, $\ket{\ket{\Phi_{\rm d}}}$ have rather different structures. 

The physical states that can be prepared via simultaneous measurements of $\hat{{\rm S}}_{\rm z}$ and $\hat{{\rm S}}^2$ in the smeared space theory are, therefore, 
%\begin{widetext}
\begin{subequations}
\begin{eqnarray} \label{Physical_State-1}
\ket{\ket{\Psi_{1}}} &=& \ket{\ket{\hat{{\rm S}}_{\rm z} = +(\hbar+\beta),\hat{{\rm S}}^2 = 2(\hbar+\beta)^2}}
\nonumber\\
&=& \alpha_{\rm 1a} \ket{\ket{\Psi_{\rm 1a}}} + \alpha_{\rm 1b} \ket{\ket{\Psi_{\rm 1b}}} + \alpha_{\rm 1c} \ket{\ket{\Psi_{\rm 1c}}} + \alpha_{\rm 1d} \ket{\ket{\Psi_{\rm 1d}}} \, ,
\nonumber\\
\end{eqnarray} 
\begin{eqnarray} \label{Physical_State-2}
\ket{\ket{\Psi_{2}}} &=& \ket{\ket{\hat{{\rm S}}_{\rm z} = -(\hbar+\beta),\hat{{\rm S}}^2 = 2(\hbar+\beta)^2}}
\nonumber\\
&=& \alpha_{\rm 2a} \ket{\ket{\Psi_{\rm 2a}}} + \alpha_{\rm 2b} \ket{\ket{\Psi_{\rm 2b}}} + \alpha_{\rm 2c} \ket{\ket{\Psi_{\rm 2c}}} + \alpha_{\rm 2d} \ket{\ket{\Psi_{\rm 2d}}} \, ,
\nonumber\\
\end{eqnarray} 
\begin{eqnarray} \label{Physical_State-3}
\ket{\ket{\Psi_{3}}} &=& \ket{\ket{\hat{{\rm S}}_{\rm z} = 0,\hat{{\rm S}}^2 = 2(\hbar+\beta)^2}}
\nonumber\\
&=& \alpha_{\rm 3a}\ket{ \ket{\Psi_{\rm 3a}}} + \alpha_{\rm 3b} \ket{\ket{\Psi_{\rm 3b}}} + \alpha_{\rm 3c} \ket{\ket{\Psi_{\rm 3c}} }+ \alpha_{\rm 3d} \ket{\ket{\Psi_{\rm 3d}}} \, ,
\nonumber\\
\end{eqnarray} 
\begin{eqnarray} \label{Physical_State-4}
\ket{\ket{\Phi}} &=& \ket{\ket{\hat{{\rm S}}_{\rm z} = 0,\hat{{\rm S}}^2 = 0}}
\nonumber\\
&=& \tilde{\alpha}_{\rm a} \ket{\ket{\Phi_{\rm a}} }+ \tilde{\alpha}_{\rm b} \ket{\ket{\Phi_{\rm b}}} + \tilde{\alpha}_{\rm c} \ket{\ket{\Phi_{\rm c}}} + \tilde{\alpha}_{\rm d} \ket{\ket{\Phi_{\rm d}}} \, ,
\nonumber\\
\end{eqnarray} 
\end{subequations}
%\end{widetext}
where  the conditions
\begin{eqnarray} \label{conditions}
\sum_{k} |\alpha_{ik}|^2 = 1 \, , \quad \sum_{k} |\tilde{\alpha}_{k}|^2 = 1 \, , 
\end{eqnarray} 
hold for all $k \in \rm \left\{a,b,c,d\right\}$ and $i \in \left\{1,2,3\right\}$.

Finally, we construct the smeared two-particle creation and annihilation operators as 
\begin{eqnarray} \label{}
\hat{{\rm S}}_{\pm} = \hat{{\rm S}}_{\rm x} \pm i \, \hat{{\rm S}}_{\rm y} = \hat{{\rm S}}_{\pm}^{\rm (A)} + \hat{{\rm S}}_{\pm}^{\rm (B)} \, . 
\end{eqnarray} 
Together with the total $z$-spin of the two-particle state, $\hat{{\rm S}}_{\rm z} = \hat{{\rm S}}_{\rm z}^{\rm (A)} + \hat{{\rm S}}_{\rm z}^{\rm (B)}$, these satisfy the algebra Eqs. (\ref{S_pm_Lie_algebra}). 
Their explicit forms are given in Eqs. (\ref{S_plus_AB_explicit})-(\ref{S_minus_AB_explicit}) and we note that, unlike their canonical counterparts Eqs. (\ref{s_pm_explicit*}), both operators are asymmetric along the main right-to-left diagonals of their respective matrices.  

The simultaneous eigenvectors of $\hat{{\rm S}}_{+}$ and $\hat{{\rm S}}^2$ are $\ket{\ket{\Psi_{\rm 1a}}}$ to $\ket{\ket{\Psi_{\rm 1d}}}$ (\ref{S_i^(AB)_eigenvector-1a})-(\ref{S_i^(AB)_eigenvector-1d}) and $\ket{\ket{\Phi_{\rm a}}}$ to $\ket{\ket{\Phi_{\rm d}}}$ (\ref{S_i^(AB)_eigenvector-4a})-(\ref{S_i^(AB)_eigenvector-4d}), plus 8 copies of the sixteen-dimensional null vector, whereas the eigenvectors of $\hat{{\rm S}}_{-}$ and $\hat{{\rm S}}^2$ are $\ket{\ket{\Psi_{\rm 2a}}}$ to $\ket{\ket{\Psi_{\rm 2d}}}$ (\ref{S_i^(AB)_eigenvector-2a})-(\ref{S_i^(AB)_eigenvector-2d}) and $\ket{\ket{\Phi_{\rm a}}}$ to $\ket{\ket{\Phi_{\rm d}}}$, plus 8 copies of the null vector. 
All 16 eigenvectors of both operators correspond to the eigenvalue 0. 
These may be compared with the eigenvectors of their canonical two-particle counterparts, Eqs. (\ref{s_pm_explicit*}). 

Using Eqs. (\ref{spin_flip_z})-(\ref{spin_flip_z'}), it is straightforward to determine the two-particle spin flips induced by (\ref{S_plus_AB_explicit})-(\ref{S_minus_AB_explicit}). 
Below, we give only a sample, in order to illustrate their main differences with the corresponding canonical expressions:
%\begin{widetext}
\begin{subequations} \label{smeared_spin_flips_ALL}
\begin{eqnarray} \label{smeared_spin_flips_S_minus}
\hat{S}_{-} \ket{\ket{ \uparrow_{\rm z} }}_{\rm A}  \ket{\ket{ \uparrow_{\rm z} }}_{\rm B} &=& \sqrt{1+\delta} \, (\hbar + i\sqrt{\hbar\beta}) \, (\ket{\ket{ \downarrow'_{\rm z} }}_{\rm A}  \ket{\ket{ \uparrow_{\rm z} }}_{\rm B}  + \ket{\ket{ \uparrow_{\rm z} }}_{\rm A}  \ket{\ket{ \downarrow'_{\rm z} }}_{\rm B} ) \, , 
\nonumber\\
\hat{S}_{-} \ket{\ket{ \uparrow'_{\rm z} }}_{\rm A}  \ket{\ket{ \uparrow_{\rm z} }}_{\rm B} &=& \sqrt{1+\delta} \, [(\hbar - i\sqrt{\hbar\beta}) \ket{\ket{ \downarrow_{\rm z} }}_{\rm A}  \ket{\ket{ \uparrow_{\rm z} }}_{\rm B}  + (\hbar + i\sqrt{\hbar\beta}) \ket{\ket{ \uparrow'_{\rm z} }}_{\rm A}  \ket{\ket{ \downarrow'_{\rm z} }}_{\rm B} ] \, , 
\nonumber\\
\hat{S}_{-} \ket{\ket{ \uparrow_{\rm z} }}_{\rm A}  \ket{\ket{ \uparrow'_{\rm z} }}_{\rm B} &=& \sqrt{1+\delta} \, [(\hbar + i\sqrt{\hbar\beta}) \ket{\ket{ \downarrow'_{\rm z} }}_{\rm A}  \ket{\ket{ \uparrow'_{\rm z} }}_{\rm B}  + (\hbar - i\sqrt{\hbar\beta}) \ket{\ket{ \uparrow_{\rm z} }}_{\rm A}  \ket{\ket{ \downarrow_{\rm z} }}_{\rm B} ] 
\nonumber\\
\hat{S}_{-} \ket{\ket{ \uparrow'_{\rm z} }}_{\rm A}  \ket{\ket{ \uparrow'_{\rm z} }}_{\rm B} &=& \sqrt{1+\delta} \, (\hbar - i\sqrt{\hbar\beta}) \, (\ket{\ket{ \downarrow_{\rm z} }}_{\rm A}  \ket{\ket{ \uparrow'_{\rm z} }}_{\rm B}  + \ket{\ket{ \uparrow'_{\rm z} }}_{\rm A}  \ket{\ket{ \downarrow_{\rm z} }}_{\rm B}) \, , 
\end{eqnarray}
\begin{eqnarray} \label{smeared_spin_flips_S_plus}
\hat{S}_{+} \ket{\ket{ \downarrow_{\rm z} }}_{\rm A}  \ket{\ket{ \downarrow_{\rm z} }}_{\rm B} &=& \sqrt{1+\delta} \, (\hbar + i\sqrt{\hbar\beta}) \, (\ket{\ket{ \uparrow'_{\rm z} }}_{\rm A}  \ket{\ket{ \downarrow_{\rm z} }}_{\rm B}  + \ket{\ket{ \downarrow_{\rm z} }}_{\rm A}  \ket{\ket{ \uparrow'_{\rm z} }}_{\rm B} ) \, , 
\nonumber\\
\hat{S}_{+} \ket{\ket{ \downarrow'_{\rm z} }}_{\rm A}  \ket{\ket{ \downarrow_{\rm z} }}_{\rm B} &=& \sqrt{1+\delta} \, [(\hbar - i\sqrt{\hbar\beta}) \ket{\ket{ \uparrow_{\rm z} }}_{\rm A}  \ket{\ket{ \downarrow_{\rm z} }}_{\rm B}  + (\hbar + i\sqrt{\hbar\beta}) \ket{\ket{ \downarrow'_{\rm z} }}_{\rm A}  \ket{\ket{ \uparrow'_{\rm z} }}_{\rm B} ] \, , 
\nonumber\\
\hat{S}_{+} \ket{\ket{ \downarrow_{\rm z} }}_{\rm A}  \ket{\ket{ \downarrow'_{\rm z} }}_{\rm B} &=& \sqrt{1+\delta} \, [(\hbar + i\sqrt{\hbar\beta}) \ket{\ket{ \uparrow'_{\rm z} }}_{\rm A}  \ket{\ket{ \downarrow'_{\rm z} }}_{\rm B}  + (\hbar - i\sqrt{\hbar\beta}) \ket{\ket{ \downarrow_{\rm z} }}_{\rm A}  \ket{\ket{ \uparrow_{\rm z} }}_{\rm B} ] 
\nonumber\\
\hat{S}_{+} \ket{\ket{ \downarrow'_{\rm z} }}_{\rm A}  \ket{\ket{ \downarrow'_{\rm z} }}_{\rm B} &=& \sqrt{1+\delta} \, (\hbar - i\sqrt{\hbar\beta}) \, (\ket{\ket{ \uparrow_{\rm z} }}_{\rm A}  \ket{\ket{ \downarrow'_{\rm z} }}_{\rm B}  + \ket{\ket{ \downarrow'_{\rm z} }}_{\rm A}  \ket{\ket{ \uparrow_{\rm z} }}_{\rm B}) \, . 
\end{eqnarray}
\end{subequations}
Equations (\ref{smeared_spin_flips_S_minus}) and (\ref{smeared_spin_flips_S_plus}) may be compared with the expressions on the top line of Eqs. (\ref{spin_flip_z-*}) and on the bottom line of Eqs. (\ref{spin_flip_z+*}), respectively. 
%\end{widetext}

%%%%%%%%%%%%%%%%%%%%%%%%%%%%%%%%%%%%%%%%%%%%%%%%%%%%%%%%%%
\subsection{Bell states in smeared space} \label{Sec.3.3}

Finally, we see that the smeared space Bell states, for spin measurements in the $i^{\rm th}$ direction, can be constructed by analogy with Eqs. (\ref{Psi_+})-(\ref{Phi_-}) as
%\begin{widetext}
\begin{subequations} \label{Smeared_Bell_states*}
\begin{eqnarray} \label{Psi_+}
\ket{\ket{\Psi_{+}^{(i)}}} = \ket{\ket{\Psi_3^{(i)}}} = \ket{\ket{\hat{{\rm S}}_{i}=0,\hat{{\rm S}}^2=2(\hbar+\beta)^2}} \, , 
\end{eqnarray}
\begin{eqnarray} \label{Psi_-}
\ket{\ket{\Psi_{-}^{(i)}}} &=& \ket{\ket{\Phi^{(i)}}} = \ket{\ket{\hat{{\rm S}}_{i}=0,\hat{{\rm S}}^2=0}} \, , 
\end{eqnarray}
\begin{eqnarray} \label{Phi_+}
\ket{\ket{\Phi_{+}^{(i)}}} &=& \frac{1}{\sqrt{2}}(\ket{\ket{\Psi_1^{(i)}}}+\ket{\ket{\Psi_2^{(i)}}}) 
\nonumber\\ 
&=&  \frac{1}{\sqrt{2}}(\ket{\ket{\hat{{\rm S}}_{i}=+(\hbar+\beta),\hat{{\rm S}}^2=2(\hbar+\beta)^2}} + \ket{\ket{\hat{{\rm S}}_{i}=-(\hbar+\beta),\hat{{\rm S}}^2=2(\hbar+\beta)^2}}) \, , 
\nonumber\\ 
\end{eqnarray}
\begin{eqnarray} \label{Phi_-}
\ket{\ket{\Phi_{-}^{(i)}}} &=& \frac{1}{\sqrt{2}}(\ket{\ket{\Psi_1^{(i)}}}-\ket{\ket{\Psi_2^{(i)}}}) 
\nonumber\\ 
&=& \frac{1}{\sqrt{2}}(\ket{\ket{\hat{{\rm S}}_{i}=+(\hbar +\beta),\hat{{\rm S}}^2=2(\hbar+\beta)^2}} - \ket{\ket{\hat{{\rm S}}_{i}=-(\hbar+\beta),\hat{{\rm S}}^2=2(\hbar+\beta)^2}}) \, , 
\nonumber\\ 
\end{eqnarray}
\end{subequations}
where 
\begin{eqnarray} \label{Physical_State-1*}
\ket{\ket{\Psi^{(i)}_{1}}} &=& \ket{\ket{\hat{{\rm S}}_{i} = +(\hbar +\beta),\hat{{\rm S}}^2 = 2(\hbar+\beta)^2}}
\nonumber\\
&=& \alpha_{\rm 1a}\ket{ \ket{\Psi^{(i)}_{\rm 1a}}} + \alpha_{\rm 1b} \ket{\ket{\Psi^{(i)}_{\rm 1b}}} + \alpha_{\rm 1c} \ket{\ket{\Psi^{(i)}_{\rm 1c}} }+ \alpha_{\rm 1d} \ket{\ket{\Psi^{(i)}_{\rm 1d}}} \, ,
\nonumber\\
\end{eqnarray} 
%\end{widetext}
and $\sum_{k} |\alpha_{1k}|^2 = 1$ for $k \in \rm \left\{a,b,c,d \right\}$, etc. 
The simultaneous eigenstates of $\hat{{\rm S}}_{i}$ and $\hat{{\rm S}}^2$, labelled $\left\{\ket{\ket{\Psi^{(i)}_{1k}}},\ket{\ket{\Psi^{(i)}_{2k}}},\ket{\ket{\Psi^{(i)}_{3k}}},\ket{\ket{\Phi^{(i)}_{k}}}\right\}$, take forms analogous to the simultaneous eigenstates of $\hat{{\rm S}}_{z}$ and $\hat{{\rm S}}^2$, denoted simply as $\left\{\ket{ \ket{\Psi_{1k}}},\ket{ \ket{\Psi_{2k}}},\ket{ \ket{\Psi_{3k}}},\ket{ \ket{\Phi_{k}}}\right\}$ in Eqs. (\ref{S_i^(AB)_eigenvectors-1})-(\ref{S_i^(AB)_eigenvectors-4}), but with $\ket{\ket{\uparrow_{\rm z}}}$ and $\ket{\ket{\downarrow_{\rm z}}}$ replaced by $\ket{\ket{\uparrow_{i}}}$ and $\ket{\ket{\downarrow_{i}}}$, for any $i \in \rm \left\{x,y,z\right\}$. 

Written explicitly, the smeared Bell states are considerably more complex than their canonical counterparts. 
Nonetheless, they remain indistinguishable from them, in {\it individual} measurements, except for a small rescaling of their eigenvalues such that $\hbar \rightarrow \hbar + \beta$. 
Empirically, this shift is undetectable, except via the existence of the spin GURs (\ref{spin_GURs})-(\ref{DS^2*}), whose middle terms contain non-canonical contributions that depend only on $\hbar$ or $\beta$, alone. 

Our analysis suggests that, if geometry is truly nonlocal at the Planck scale, we must look for signatures of entanglement between material particles and the quantum spatial background in the statistical correlations between subspaces of multiparticle states. 
Such non-canonical correlations are a direct result of the ${\rm SU(2)}$ invariance of the composite wave vector $\ket{\ket{\Psi}}_{\rm spin}$ (\ref{pre-measurement_Psi_spin}). 
We repeat that this corresponds to the state of a spinning particle, interacting with the nonlocal background geometry, as opposed to the state of a spinning particle propagating in classical space. 

The representation of ${\rm SU(2)}$ generated by the generalised spin matrices, Eqs. (\ref{S_i_explicit}), describes the `internal' gauge symmetry that is compatible with the EGUP (\ref{EGUP-2}), and with the corresponding smearing of the `external' symmetry algebra, ${\rm SO(3)}$, that it implies. 
This representation is characterised by the subalgebras (\ref{rearrange-S.1})-(\ref{rearrange-S.5}) and (\ref{Clifford-S.1})-(\ref{Clifford-S.5}), that also give rise to spin GURs. 
If our model is correct, these are the `smoking gun' of nonlocal geometry in the low-energy regime of the spin sector.

%%%%%%%%%%%%%%%%%%%%%%%%%%%%%%%%%%%%%%%%%%%%%%%%%%%%%%%%%%
%%%%%%%%%%%%%%%%%%%%%%%%%%%%%%%%%%%%%%%%%%%%%%%%%%%%%%%%%%
\section{The smeared-space representation of {\rm SU(2)}} \label{Sec.4}

In the canonical representation, a general element of ${\rm SU(2)}$ is written as
\begin{eqnarray} \label{U(x)}
U = u_0\mathbb{I} + i \, \bold{u} \, . \, \vec{\sigma} = 
\left[
\begin{matrix} 
u_0 + iu_3 && u_2 + iu_1 \\
-u_2 + iu_1 && u_0 - iu_3 
\end{matrix}
\right] \, , 
\end{eqnarray}
where $\bold{u} = (u_1,u_2,u_3)$, with $u_0 \, , u_i \in \mathbb{R}$, $\vec{\sigma} = (\sigma_1,\sigma_2,\sigma_3)$, 
and
\begin{eqnarray} \label{detU}
{\rm det} \, U = u_0^2 + u_1^2 + u_2^2 + u_3^2 = 1 \, . 
\end{eqnarray}
The fundamental relation, $\sigma_i \sigma_j = \delta_{ij} \, \mathbb{I} + i \, \epsilon_{ij}{}^{k}\sigma_{k}$, ensures that the product of two matrices of the form (\ref{U(x)}) is again of the same form. 
This shows that ordinary matrix multiplication is a valid group composition law. 
Equation (\ref{detU}) defines the unit $3$-sphere, embedded in $\mathbb{R}^4$, so that
\begin{eqnarray} \label{SU(2)=S^3}
{\rm SU(2)} \cong {\rm S}^3 \, . 
\end{eqnarray}

It is also useful to parameterise the elements of ${\rm SU(2)}$ in a different way, in terms of the components of a unit vector,  
\begin{eqnarray} \label{n_vec}
\bold{n} = -\bold{u}/|\bold{u}| \, ,
\end{eqnarray}
and an angle $\theta$, such that 
\begin{eqnarray} \label{theta_param}
\cos(\theta/2) = u_0 \, , \quad -\sin(\theta/2)n_i = u_i \, . 
\end{eqnarray}
These identifications ensure that $0 \leq |\bold{u}| \leq 1$ for $\theta \in [0,2\pi]$. 
We then have $U \equiv U_{\bold{n}}(\theta)$, where
\begin{eqnarray} \label{U(theta)}
U_{\bold{n}}(\theta) = \cos(\theta/2) \mathbb{I} - i \sin(\theta/2) \bold{n} \, . \, \vec{\sigma} = \exp\left(-\frac{i\theta \, \bold{n} \, . \, \vec{\sigma}}{2}\right) \, . 
\end{eqnarray}

Next, we note that the matrices $\left\{\Sigma_i \right\}_{i=1}^{3}$, defined as
\begin{eqnarray} \label{Sigma_i}
\Sigma_i := \frac{2}{\hbar + \beta} \hat{\rm S}_i = \frac{1}{1+\delta}(\sigma_{i} \otimes \mathbb{I} + \delta \, \mathbb{I} \otimes \sigma_{i} + \sqrt{\delta} \, \epsilon_{i}{}^{jk}\sigma_{j} \otimes \sigma_{k}) \, , 
\end{eqnarray}
satisfy the algebras
\begin{eqnarray} \label{Sigma_i_algebras}
[\Sigma_i,\Sigma_j] = 2i \, \epsilon_{ij}{}^{k} \Sigma_k \, , \quad \left\{\Sigma_i,\Sigma_j\right\} = 2\delta_{ij} \, \mathbb{I} \, , 
\end{eqnarray}
which, together, are equivalent to the fundamental relation
\begin{eqnarray} \label{Sigma_i_fund_rel}
\Sigma_i \Sigma_j = \delta_{ij} \, \mathbb{I} + i\epsilon_{ij}{}^{k} \Sigma_k \, . 
\end{eqnarray}
These are the smeared space analogues of the canonical Pauli matrices, $\sigma_i = (2/\hbar)\hat{s}_i$. 
Unlike $\left\{\sigma_i \right\}_{i=1}^{3}$, whose matrix elements do not depend on any parameters, or constants, of a physical theory, $\left\{\Sigma_i \right\}_{i=1}^{3}$ depend explicitly on the smearing parameter, $\delta = \hbar/\beta$ (\ref{delta}). 
Another important difference between the smeared-space sigma matrices and the canonical Pauli matrices is that
\begin{eqnarray} \label{}
{\rm tr} \, \Sigma_i = 0 \, , \quad {\rm det} \, \Sigma_i = 1 \, ,  
\end{eqnarray}
whereas ${\rm tr} \, \sigma_i = 0$ and ${\rm det} \, \sigma_i = -1$. 
The $\left\{\Sigma_i \right\}_{i=1}^{3}$ matrices are hermitian, and also unitary, since
\begin{eqnarray} \label{}
\Sigma_i = \Sigma_i^{\dagger} = \Sigma_i^{-1} = (\Sigma_i^{-1}){}^{\dagger} \, , 
\end{eqnarray}
which is exactly analogous to the relation $\sigma_i = \sigma_i^{\dagger} = \sigma_i^{-1} = (\sigma_i^{-1} ){}^{\dagger}$. 

Exponentiating the $\Sigma_i$ generators gives
\begin{eqnarray} \label{mathcal_U(theta)}
\mathfrak{U}_{\bold{n}}(\theta) = \cos(\theta/2) \mathbb{I} - i \sin(\theta/2) \bold{n} \, . \, \vec{\Sigma} = \exp\left(-\frac{i\theta \, \bold{n} \, . \, \vec{\Sigma}}{2}\right) \, ,  
\end{eqnarray}
or, equivalently, 
\begin{eqnarray} \label{mathcal_U(x)}
\mathfrak{U} = u_0\mathbb{I} + i \, \bold{u} \, . \, \vec{\Sigma} = %\bold{\sigma}
\left[\begin{matrix}
u_0 + iu_3 && \frac{(u_2+iu_1)\sqrt{\delta}}{-i + \sqrt{\delta}} && \frac{u1-iu_2}{-i + \sqrt{\delta}} && 0 \\
\frac{(-u_2+iu_1)\sqrt{\delta}}{i + \sqrt{\delta}} && u_0 + i\left(\frac{1-\delta}{1+\delta}\right)u_3 && - \frac{2u_3\sqrt{\delta}}{1+\delta} && -\frac{u_1-iu_2}{i+\sqrt{\delta}} \\
-\frac{u_1+iu_2}{i+\sqrt{\delta}} && \frac{2u_3\sqrt{\delta}}{1+\delta} && u_0 - i\left(\frac{1-\delta}{1+\delta}\right)u_3 && \frac{(u_2+iu_1)\sqrt{\delta}}{i + \sqrt{\delta}} \\
0 && \frac{u_1+iu_2}{-i+\sqrt{\delta}} && \frac{(-u_2+iu_1)\sqrt{\delta}}{-i + \sqrt{\delta}} && u_0 - iu_3
\end{matrix}\right] \, , 
\end{eqnarray}
where $\vec{\Sigma} = (\Sigma_1,\Sigma_2,\Sigma_3)$ and we have again used Eq. (\ref{theta_param}). 
We then have
\begin{eqnarray} \label{mathcal_U(x)_tr_det}
{\rm tr} \, \mathfrak{U} = 4u_0 \, , \quad {\rm det} \, \mathfrak{U} = (u_0^2+u_1^2+u_2^2+u_3^2)^2 \, ,
\end{eqnarray}
which may be compared with the equivalent expressions for $U$ (\ref{U(x)}), i.e., ${\rm tr} \, U = 2u_0$ and ${\rm det} \, U = u_0^2+u_1^2+u_2^2+u_3^2$ (\ref{detU}). 

Imposing ${\rm det} \, \mathfrak{U} = 1$ is then equivalent to imposing  
\begin{eqnarray} \label{det_mathcal_U=1}
u_0^2+u_1^2+u_2^2+u_3^2 = \pm 1 \, .
\end{eqnarray} 
However, since $u_{\mu} \in \mathbb{R}$ for all $\mu \in \left\{0,1,2,3\right\}$, this condition can only be satisfied by the first of Eqs. (\ref{det_mathcal_U=1}), which corresponds to choosing $+1$ on the right-hand side. 
This demonstrates that $\left\{\Sigma_i\right\}_{i=1}^{3}$ generates a non-canonical representation of the canonical spin group, with $\mathfrak{U} \cong U \in {\rm SU(2)} \cong {\rm S^3}$ for a given set of parameters, $\left\{u_{\mu}\right\}_{\mu=0}^{3}$. 

In other words, the smeared-space spin group is simply the canonical spin group, ${\rm SU(2)}$, but $\mathfrak{U}$ (\ref{mathcal_U(x)}) is distinct from the `trivial' four-dimensional representations, $U \otimes \mathbb{I}$ and $\mathbb{I} \otimes U$, where $U$ is given by Eq. (\ref{U(x)}). 
We refer to these representations as `trivial' because they act nontrivially only on one subspace of the tensor product and trivially on the remainder of the composite state. 
In stricter terminology, they are spin-$1/2$ representations that are trivially embedded in a higher-dimensional space, not trivial spin-$0$ representations of the kind considered in formal group theory. 
Their generators are $\sigma_i \otimes \mathbb{I}$ and $\mathbb{I} \otimes \sigma_i$, respectively, which are obtained as different limits of the general expression for $\Sigma_i$ (\ref{Sigma_i}). 
The former corresponds to the unsmeared limit, $\beta \rightarrow 0$, and the latter is obtained by taking $\hbar \rightarrow 0$ and $\beta \rightarrow \hbar$. 
Each limit corresponds to one particle of a two-particle state in canonical quantum mechanics. 
The $\Sigma_i$ matrices are also inequivalent to the four-dimensional generators of ${\rm SU(2) \times SU(2)}$, which is the internal symmetry group of the canonical two-fermion state. 
Its generators are written as the sum $\sigma_i \otimes \mathbb{I} + \mathbb{I} \otimes \sigma_i$, which emerges from Eq. (\ref{Sigma_i}) when the interaction term, $2(\hbar + \beta)^{-1}\hat{\mathbb{S}}_{i}$, is ignored, and $\beta \rightarrow \hbar$. 

In most models of quantum geometry including matter, it is conventionally assumed that spin-$1/2$ SU(2) symmetry applies only to canonical fermions in the matter sector, rather than to the composite state that incorporates matter-geometry interactions. 
In this case, the spin operators for single fermions take the simple form $\hat{s}_{i} = (\hbar/2)(\sigma_i \otimes \mathbb{I})$ (*), where the identity acts on the geometric part. 
The representation (\ref{mathcal_U(x)}) is different in that it seeks to describe the SU(2) invariance of fermionic matter interacting with a fluctuating quantum spacetime. 
The corresponding spin operators are not equivalent to (*), but act nontrivially on both subspaces of the tensor product state. 
Furthermore, the noncanonical effects of this representation are determined by the smearing parameter $\delta = \hbar/\beta \simeq \hbar \times 10^{-61}$, previously employed to derive the EGUP (\ref{EGUP-2}), and the predictions of canonical models are recovered in the limit $\delta \rightarrow 0$. 

This construction does not imply that the total quantum state of the background is invariant under a spin-$1/2$ representation SU(2), but only that the composite state describing its interactions with the canonical fermions should be. 
Theoretically, the force-mediating bosons that transmit spacetime fluctuations to the matter sector may have any spin, although it widely accepted that gravitons should have spin-2 \cite{Pauli-Fierz}. 
In \cite{Lake:2019nmn,Lake:2020rwc}, it was argued that a distinction should be drawn between the quanta of gravity, i.e., spacetime curvature, and the quanta of spacetime itself. 
This raises the intriguing possibility that the latter could still be fermionic, with spin $\pm \beta/2$, but our model must be extended to the relativistic regime before this issue can be clarified.  

%%%%%%%%%%%%%%%%%%%%%%%%%%%%%%%%%%%%%%%%%%%%%%%%%%%%%%%%%%
%%%%%%%%%%%%%%%%%%%%%%%%%%%%%%%%%%%%%%%%%%%%%%%%%%%%%%%%%%
\section{Discussion} \label{Sec.5}

%%%%%%%%%%%%%%%%%%%%%%%%%%%%%%%%%%%%%%%%%%%%%%%%%%%%%%%%%%
\subsection{Conclusions} \label{Sec.5.1}

We have proposed a new model of {\it quantum} nonlocal geometry \cite{Lake:2018zeg,Lake:2019nmn,Lake:2020chb,Lake:2020rwc} and have investigated its consequences for both external and internal symmetries. 
(Though many models of nonlocal geometry have been proposed in the literature, most are intrinsically {\it classical} in nature. See \cite{Lake:2020rwc} for further discussion of this point). 
In the position space representation, classical spatial points are `smeared' over the Planck volume, whereas, in the momentum space representation, momentum space points are smeared over the volume associated with the de Sitter mass. 
The smearing of the canonical phase space generates generalised uncertainty relations (GURs), including the generalised uncertainty principle (GUP), extended uncertainty principle (EUP), and extended generalised uncertainty principle (EGUP), previously proposed in the quantum gravity literature \cite{Adler:1999bu,Scardigli:1999jh,Bolen:2004sq,Park:2007az,Bambi:2007ty}, as well as GURs for angular momentum. 
This, in turn, suggests an analogous generalisation of the spin uncertainty relations. 

We have shown that the spin GURs can be obtained from set of generalised spin-measurement operators, $\hat{{\rm S}}_{i}$, each of which is given by the sum of three subcomponents. 
The first subcomponent, $\hat{\mathcal{S}}_{i}$, acts nontrivially only on the Hilbert space of the canonical quantum particle(s), whereas the second, $\hat{\mathcal{S}}'_{i}$, acts nontrivially only on the Hilbert space that describes their interactions with the fluctuating background geometry. 
The third, $\hat{\mathbb{S}}_{i}$, acts nontrivially on both subspaces and can also be interpreted as a noncanonical interaction term \cite{Lake:2019nmn,Lake:2020rwc}. 

The subcomponents $\left\{\hat{\mathcal{S}}_{i},\hat{\mathcal{S}}'_{i} ,\hat{\mathbb{S}}_{i}\right\}_{i=1}^{3}$ obey the generalised Lie and Clifford algebras, Eqs. (\ref{rearrange-S.1})-(\ref{rearrange-S.5}) and (\ref{Clifford-S.1})-(\ref{Clifford-S.5}), and the 
$\hat{{\rm S}}_{i}$ operators for one-particle states are four-dimensional, possessing 4 independent eigenvectors. 
For the $\hat{{\rm S}}_{\rm z}$ operator, these can be split into two spin `up' states, $\left\{\ket{\ket{\uparrow_{\rm z}}}, \ket{\ket{\uparrow'_{\rm z}}}\right\}$, corresponding to the eigenvalue $+(\hbar + \beta)/2$, and two spin `down' states, $\left\{\ket{\ket{\downarrow_{\rm z}}}, \ket{\ket{\downarrow'_{\rm z}}}\right\}$, corresponding to $-(\hbar + \beta)/2$, where $\beta \simeq \hbar \times 10^{-61}$ is the quantum of action for the background geometry \cite{Lake:2018zeg,Lake:2019nmn,Lake:2020chb,Lake:2020rwc}. 
The double ket notation indicates that the eigenvectors exist in the tensor product Hilbert space of the composite state describing matter-geometry interactions. 
Crucially, we found that the unprimed eigenvectors can be written as simple product states, whereas the primed eigenvectors represent states in which the spin of the particle is entangled with the spin sector of the geometry. 

Finally, we extended our analysis to smeared two-particle states, focussing on the simultaneous eigenvectors of $\hat{{\rm S}}_{\rm z}$ and $\hat{{\rm S}}^2$, and on the two-particle creation and annihilation operators $\hat{{\rm S}}_{\pm}$. 
We have shown that, for every eigenvector of the corresponding operator in canonical quantum theory, there exist 4 eigenvectors in smeared space. 
These take radically different forms from their canonical counterparts, but, remarkably, the corresponding eigenvalues differ only by the simple rescaling $\hbar \rightarrow \hbar + \beta$. 
For this reason, it is very unlikely that smeared states can be distinguished from unsmeared states, using {\it individual} measurements, with current or near-future technology. 
Nonetheless, the existence of spin GURs provides a distinct experimental signature of the model and it may be hoped that the nonlocality of the quantum geometry can be probed, indirectly, by measuring the noncanonical correlations between material particles, induced by their interaction with the background. 
In a future work, we will investigate this possibility in more detail. 
Below, we outline several other avenues for future research. 

%%%%%%%%%%%%%%%%%%%%%%%%%%%%%%%%%%%%%%%%%%%%%%%%%%%%%%%%%%
\subsection{Future work} \label{Sec.5.2}

There is still lots of work to be done. 
In addition to further studies of the spin GURs, for example, constructing the smeared space generalisations of the canonical Bell and CHSH inequalities, the model has immediate implications for the information loss paradox \cite{Chow:2020iyn,Almheiri:2020cfm}, relativity, and particle physics. 
These include the following:

\begin{itemize}

\item In \cite{Lake:2018zeg}, we argued that a smeared spatial `point' can be viewed as a quantum reference frame (QRF). 
However, there is a fundamental difference between a QRF embedded as a material quantum system in a classical background geometry, as proposed in \cite{Giacomini:2017zju}, and one generated by the nonlocality of the geometry itself  \cite{Lake:2018zeg,Lake:2020rwc}. 
Realistic observers are embedded as material quantum systems in {\it quantum} geometries, so that the smeared-space and canonical QRF formalisms should be combined to describe this scenario.  

\item The model should be extended to the relativistic regime via an appropriate smearing of the time coordinate and we expect this to generate a generalisation of the canonical Poincar{\' e} algebra. 
The generalisation should support an appropriate subalgebra structure that classifies the symmetries of nonlocal Minkowski space. 
Generalisations of the canonical Dirac and Klein-Gordon equations can then be obtained, which are compatible with the minimum length and minimum momentum phenomenology implied by the EGUP.

\item If the subalgebra structure of the Poincar{\' e} group, which classifies the symmetries of nonlocal spacetime, can be consistently constructed, this has immediate consequences for the Standard Model of particle physics. 
The seminal realisation, by Eugene Wigner, that canonical quantum particles are ``unitary representations of the inhomogeneous Lorentz group'' \cite{Wigner:1939cj}, is conventionally applied to particles propagating in classical spacetimes. 
We aim to construct representations that act nontrivially on both subspaces of a tensor product Hilbert space describing matter-geometry interactions. 
From these representations, we can expect to obtain minimum length and minimum momentum phenomenology that remains consistent with the symmetries of special relativity.  

\item Finally, the existence of qubits that are entangled with the spacetime background may have profound implications for the black hole information loss paradox \cite{Chow:2020iyn}. 
Consider, for example, two black holes with identical mass, charge, and `spin'. 
(In the common terminology, the `spin' of a black hole refers to its orbital angular momentum \cite{Chandrasekhar:1985kt} and must not be confused with genuine quantum mechanical spin.) 
Let us imagine that the first black hole emits a particle of Hawking radiation, say, an electron, in the spin `up' eigenstate $\ket{\ket{\uparrow_{\rm z}}}$, whereas the second emits an electron in the spin `up' eigenstate $\ket{\ket{\uparrow'_{\rm z}}}$. 
Due to recoil, both particles are entangled with their respective black holes, and, according to the smeared space theory, the states of the two black-hole-plus-electron systems are indistinguishable, via simultaneous measurements of $\hat{{\rm S}}_{\rm z}$ and $\hat{{\rm S}}^2$. 
Nonetheless, the electron in state $\ket{\ket{\uparrow'_{\rm z}}}$ possesses {\it additional} entanglement, with the nonlocal spacetime background, and, hence, additional entanglement entropy. 
Such `geometric' qubits may radiate more entropy away from the black hole than their canonical counterparts, and, therefore, more information. 
According to our previous arguments, this information should be encoded in additional {\it noncanonical} correlations between the subsystems of multiparticle states, due to their mutual interaction with the nonlocal background geometry. 
However, existing models of black hole evaporation do not account for this type of entanglement \cite{Chow:2020iyn}, which is explicitly generated by the delocalisation of spatial `points' over regions comparable to the Planck volume.  

\end{itemize}

%\vspace{6pt} 

%Acknowledgements%%%%%%%%%%%%%%%%%%%%%%%%%%%%%%%%%%%%%%%%%%%%%
\section*{Acknowledgements}
ML is supported by a grant from the Guangdong Natural Science Foundation. Details are available at 
${\rm http://gdstc.gd.gov.cn/zwgk\_n/tzgg/content/post\_3123839.html}$.

\appendix

%Appendix%%%%%%%%%%%%%%%%%%%%%%%%%%%%%%%%%%%%%%%%%%%%%%%%%%%%
%%%%%%%%%%%%%%%%%%%%%%%%%%%%%%%%%%%%%%%%%%%%%%%%%%%%%%%%%%
\section{Spin in canonical quantum mechanics} \label{Appendix-A}

In this Appendix, we briefly review the treatment of one- and two-particle states in canonical quantum mechanics. 
The structure of Secs. (\ref{Sec.A.1})-(\ref{Sec.A.3}) parallels that of Secs. (\ref{Sec.3.1})-(\ref{Sec.3.3}), so that the results of the smeared space theory can be easily compared and contrasted with their canonical counterparts. 

%%%%%%%%%%%%%%%%%%%%%%%%%%%%%%%%%%%%%%%%%%%%%%%%%%%%%%%%%%
\subsection{One-particle systems} \label{Sec.A.1}

The spin-1/2 Pauli matrices are
\begin{eqnarray} \label{Pauli_matrices}
\sigma_{\rm x} = 
\begin{bmatrix}
    0  &  1 \\
    1  &  0 
\end{bmatrix}
\, , \quad  
\sigma_{\rm y} = 
\begin{bmatrix}
    0  &  -i \\
    i  &  0 
\end{bmatrix}
\, , \quad  
\sigma_{\rm z} = 
\begin{bmatrix}
    1  &  0 \\
    0  &  -1 
\end{bmatrix}
\, .
\end{eqnarray}
For a one-particle state, the spin-measurement operators are 
\begin{eqnarray} \label{s^i}
\hat{s}_{i} = \frac{\hbar}{2} \sigma_{i} \, ,
\end{eqnarray}
and the total spin-squared operator is a Casimir operator, 
\begin{eqnarray} \label{s^2}
\hat{s}^2 = \sum_{i=1}^{3} \hat{s}_{i}{}^2 = \frac{3\hbar^2}{4} \mathbb{I} \, ,
\end{eqnarray}
where $\mathbb{I}$ denotes the two-dimensional identity matrix. 
These satisfy the canonical Lie algebra
\begin{subequations} \label{canonical_spin_Lie_algebra}
\begin{eqnarray} \label{[s^i,s^j]}
[\hat{s}_{i},\hat{s}_{j}] = i\hbar \, \epsilon_{ij}{}^{k}\hat{s}_{k} \, ,
\end{eqnarray}
\begin{eqnarray} \label{[s^2,s^i]}
[\hat{s}_{i},\hat{s}^{2}] = 0 \, ,
\end{eqnarray}
\end{subequations}
and the canonical Clifford algebra
\begin{eqnarray} \label{[s^i,s^j]_+}
\left\{\hat{s}_{i},\hat{s}_{j}\right\} = i\frac{\hbar^2}{2} \, \delta_{ij} \mathbb{I} \, ,
\end{eqnarray}
where $\left\{ \, . \, , \, . \, \right\}$ denotes the anticommutator. %and  $i,j \in \left\{1,2,3\right\} \equiv \left\{\rm x,y,z\right\}$. 

The one-particle spin eigenvectors are
\begin{subequations} \label{s_i_eigenvectors}
\begin{eqnarray} \label{s_x_eigenvectors}
\ket{\uparrow_{\rm x}} &=& |\hat{s}_{\rm x} = +\hbar/2,\hat{s}^2 = 3\hbar^2/4\rangle = 
\frac{1}{\sqrt{2}} (1,1) \, , 
\nonumber\\
\ket{\downarrow_{\rm x}} &=& |\hat{s}_{\rm x} = -\hbar/2,\hat{s}^2 = 3\hbar^2/4\rangle = 
\frac{1}{\sqrt{2}} (1,-1) \, ,
\nonumber\\
\end{eqnarray}
\begin{eqnarray} \label{s_y_eigenvectors}
\ket{\uparrow_{\rm y}} &=& |\hat{s}_{\rm y} = +\hbar/2,\hat{s}^2 = 3\hbar^2/4\rangle = 
\frac{1}{\sqrt{2}} (1,i) \, , 
\nonumber\\
\ket{\downarrow_{\rm y}} &=& |\hat{s}_{\rm y} = -\hbar/2,\hat{s}^2 = 3\hbar^2/4\rangle = 
\frac{1}{\sqrt{2}} (1,-i) \, ,
\nonumber\\
\end{eqnarray}
\begin{eqnarray} \label{s_z_eigenvectors}
\ket{\uparrow_{\rm z}} &=& |\hat{s}_{\rm z} = +\hbar/2,\hat{s}^2 = 3\hbar^2/4\rangle = (1,0) \, , 
\nonumber\\
\ket{\downarrow_{\rm z}} &=& |\hat{s}_{\rm z} = -\hbar/2,\hat{s}^2 = 3\hbar^2/4\rangle = (0,1) \, ,
\nonumber\\
\end{eqnarray}
\end{subequations}
respectively. 
The eigenvectors of $\hat{s}_{\rm y}$ and $\hat{s}_{\rm x}$ may be written in terms of the $\hat{s}_{\rm z}$ eigenvectors as
\begin{eqnarray} \label{y_up_down}
\ket{\uparrow_{\rm y}} = \frac{1}{\sqrt{2}}(\ket{\uparrow_{\rm z}} + i\ket{\downarrow_{\rm z}}) \, , \, \, \, \ket{\downarrow_{\rm y}} = \frac{1}{\sqrt{2}}(\ket{\uparrow_{\rm z}} - i\ket{\downarrow_{\rm z}}) \, , 
\nonumber\\
\end{eqnarray}
and
\begin{eqnarray} \label{x_up_down}
\ket{\uparrow_{\rm x}} = \frac{1}{\sqrt{2}}(\ket{\uparrow_{\rm z}} + \ket{\downarrow_{\rm z}}) \, , \, \, \, \ket{\downarrow_{\rm x}} = \frac{1}{\sqrt{2}}(\ket{\uparrow_{\rm z}} - \ket{\downarrow_{\rm z}}) \, .
\nonumber\\
\end{eqnarray}
From Eqs. (\ref{[s^i,s^j]}), the single-particle spin uncertainty relations are
\begin{eqnarray} \label{[s^i,s^j]_UR}
\Delta s_{i} \, \Delta s_{j} \geq \frac{\hbar}{2} \, |\epsilon_{ij}{}^{k}\langle\hat{s}_{k}\rangle_{\psi}| \, .
\end{eqnarray}

It is also convenient to define the creation and annihilation operators, 
\begin{eqnarray} \label{sigma_pm}
\hat{s}_{\pm} = \hat{s}_{\rm x} \pm i\hat{s}_{\rm y} \, , 
\end{eqnarray}
which satisfy the algebra
\begin{eqnarray} \label{s_pm_Lie_algebra}
[\hat{s}_{\rm z},\hat{s}_{\pm}] = \pm \hbar \, \hat{s}_{\pm} \, , \quad [\hat{s}_{+},\hat{s}_{-}] = 2\hbar \, \hat{s}_{\rm z} \, .
\end{eqnarray}
Written explicitly, $\hat{s}_{\pm}$ take the form
\begin{eqnarray} \label{s_pm_explicit}
\hat{s}_{+} = \hbar
\begin{bmatrix}
    0  & 1 \\
    0  & 0  
\end{bmatrix}
\, , \quad 
\hat{s}_{-} = \hbar
\begin{bmatrix}
    0  & 0 \\
    1  & 0 
\end{bmatrix}
\, .
\end{eqnarray}
The eigenvectors of $\hat{s}_{+}$ are the null vector, $\ket{\rm null } = (0,0)$, and the $z$-spin up eigenstate, $\ket{\uparrow_{\rm z}} = (1,0)$, whereas the eigenvectors of $\hat{s}_{-}$ are the null vector and the $z$-spin down eigenstate, $
\ket{\downarrow_{\rm z}} = (0,1)$. 
For each operator, both eigenvectors correspond to the eigenvalue $0$. 
The creation and annihilation operators perform spin flips according to:
\begin{eqnarray} \label{spin_flip_z**}
\hat{s}_{-} \ket{\uparrow_{\rm z}} = \hbar \, \ket{\downarrow_{\rm z}} \, , \quad 
\hat{s}_{+} \ket{\downarrow_{\rm z}} = \hbar \, \ket{\uparrow_{\rm z}} \, . 
\end{eqnarray}
Finally, we note that the canonical spin states obey the braket relations
\begin{eqnarray} \label{canonical_spin_braket_relations}
\braket{ \uparrow_{i} | \uparrow_{i} } &=& 1 \, , \quad \braket{ \uparrow_{i} | \downarrow_{i} } = 0 \, ,
\nonumber\\
\braket{ \downarrow_{i} | \uparrow_{i} } &=& 0 \, , \quad \braket{ \downarrow_{i} | \downarrow_{i} } = 1 \, .
\end{eqnarray}

%%%%%%%%%%%%%%%%%%%%%%%%%%%%%%%%%%%%%%%%%%%%%%%%%%%%%%%%%%
\subsection{Two-particle systems} \label{Sec.A.2}

For a two-particle state, in which the particles do not interact, the spin-measurement operators are
\begin{eqnarray} \label{s^i_AB}
\hat{s}^{\rm (A)}_{i} = \frac{\hbar}{2} \, (\sigma^{\rm (A)}_{i} \otimes \mathbb{I}_{\rm B}) \, ,
\quad
\hat{s}^{\rm (B)}_{i} = \frac{\hbar}{2} \, (\mathbb{I}_{\rm A} \otimes \sigma^{\rm (B)}_{i}) \, .
\end{eqnarray}
Each set of operators $\left\{\hat{s}^{\rm (A)}_{i}\right\}_{i=1}^{3}$ and $\left\{\hat{s}^{\rm (B)}_{i}\right\}_{i=1}^{3}$ satisfies the canonical Lie algebra (\ref{[s^i,s^j]})-(\ref{[s^2,s^i]}) and the two sets commute,
\begin{eqnarray} \label{[s^i_A,s^j_B]}
[\hat{s}^{\rm (A)}_{i},\hat{s}^{\rm (B)}_{j}] = 0 \, . 
\end{eqnarray}

The total spin in the $i^{\rm th}$ direction is given by
\begin{eqnarray} \label{s_i^(AB)}
\hat{s}_{i} = \hat{s}^{\rm (A)}_{i} + \hat{s}^{\rm (B)}_{i} \, 
\end{eqnarray}
and each $\hat{s}_{i}$ has four independent eigenvectors, 
\begin{eqnarray} \label{s_z_eigenvectors}
\left\{\ket{\uparrow_{i}}_{\rm A}\ket{\uparrow_{i}}_{\rm B},\ket{\uparrow_{i}}_{\rm A}\ket{\downarrow_{i}}_{\rm B},\ket{\downarrow_{i}}_{\rm A}\ket{\uparrow_{i}}_{\rm B},\ket{\downarrow_{i}}_{\rm A}\ket{\downarrow_{i}}_{\rm B}\right\} \, . 
\nonumber\\
\end{eqnarray}
These correspond to the eigenvalues $\left\{+\hbar,0,0,-\hbar\right\}$, respectively.

The total spin-squared is
\begin{eqnarray} \label{s^2_AB}
\hat{s}^{2} = \sum_{i=1}^{3} (\hat{s}^{\rm (A)}_{i} + \hat{s}^{\rm (B)}_{i})^2 
%\nonumber\\
=  (\hat{\bold{s}}_{\rm A})^2 + 2 \, \hat{\bold{s}}_{\rm A} \, . \, \hat{\bold{s}}_{\rm B} + (\hat{\bold{s}}_{\rm B})^2 \, ,
\end{eqnarray}
where 
\begin{eqnarray} \label{s_AB_vec}
\hat{\bold{s}}_{\rm A / B} = \sum_{i=1}^{3} \hat{s}^{\rm (A / B)}_{i} \, \bold{e}_{i}(0) \, , 
\end{eqnarray}
and $\left\{\bold{e}_{i}(0)\right\}_{i=1}^{3}$ are the tangent vectors at the coordinate origin. 
From Eqs. (\ref{[s^i_A,s^j_B]}) and (\ref{s^2_AB}), we see that $\left\{\hat{s}_{i}\right\}_{i=1}^{3}$ and $\hat{s}^2$ for the two-particle state satisfy the canonical Lie algebra (\ref{[s^i,s^j]})-(\ref{[s^2,s^i]}), and the canonical Clifford algebra (\ref{[s^i,s^j]_+}), where $\mathbb{I}$ now denotes the four-dimensional identify matrix, in complete analogy with the one-particle case. 

Written explicitly, $\hat{s}_{\rm z}$ and $\hat{s}^{2}$ take the form
\begin{eqnarray} \label{s^2_AB_explicit}
\hat{s}_{\rm z} 
= \hbar
\begin{bmatrix}
    1  & 0 & 0 & 0 \\
    0  & 0 & 0 & 0 \\  
    0  & 0 & 0 & 0 \\
    0  & 0 & 0 & -1 \,
\end{bmatrix}
\, , \quad 
\hat{s}^{2} 
= \hbar^2
\begin{bmatrix}
    2  & 0 & 0 & 0 \\
    0  & 1 & 1 & 0 \\  
    0  & 1 & 1 & 0 \\
    0  & 0 & 0 & 2 \,
\end{bmatrix}
\, .
\end{eqnarray}
Note that, unlike the spin-squared operator for the one-particle state, this $\hat{s}^{2}$ is not a Casimir operator. 
Its eigenvectors are
\begin{subequations} \label{s_i^(AB)_eigenvectors}
\begin{eqnarray} \label{s_i^(AB)_eigenvector-1}
\ket{\Psi_1} &=& \ket{\uparrow_{\rm z}}_{\rm A}\ket{\uparrow_{\rm z}}_{\rm B}  
\nonumber\\
&=& \ket{\hat{s}_{\rm z} = +\hbar,\hat{s}^2 = 2\hbar^2} = (1,0,0,0) \, , 
\nonumber\\
\end{eqnarray} 
\begin{eqnarray} \label{s_i^(AB)_eigenvector-2}
\ket{\Psi_2} &=& \ket{\downarrow_{\rm z}}_{\rm A}\ket{\downarrow_{\rm z}}_{\rm B}  
\nonumber\\
&=& \ket{\hat{s}_{\rm z} = -\hbar,\hat{s}^2 = 2\hbar^2} = (0,0,0,1) \, , 
\nonumber\\
\end{eqnarray} 
\begin{eqnarray} \label{s_i^(AB)_eigenvector-3}
\ket{\Psi_3} &=& \frac{1}{\sqrt{2}}(\ket{\uparrow_{\rm z}}_{\rm A}\ket{\downarrow_{\rm z}}_{\rm B} + \ket{\downarrow_{\rm z}}_{\rm A}\ket{\uparrow_{\rm z}}_{\rm B})   
\nonumber\\
&=& \ket{\hat{s}_{\rm z} = 0,\hat{s}^2 = 2\hbar^2} = \frac{1}{\sqrt{2}} (0,1,1,0) \, , 
\nonumber\\
\end{eqnarray} 
\begin{eqnarray} \label{s_i^(AB)_eigenvector-4}
\ket{\Phi} &=& \frac{1}{\sqrt{2}}(\ket{\uparrow_{\rm z}}_{\rm A}\ket{\downarrow_{\rm z}}_{\rm B} - \ket{\downarrow_{\rm z}}_{\rm A}\ket{\uparrow_{\rm z}}_{\rm B})   
\nonumber\\
&=& \ket{\hat{s}_{\rm z} = 0,\hat{s}^2 = 0} = \frac{1}{\sqrt{2}} (0,-1,1,0) \, .
\nonumber\\
\end{eqnarray} 
\end{subequations}
These form a triplet $\left\{\ket{\Psi_1},\ket{\Psi_2},\ket{\Psi_3}\right\}$ with total spin $2\hbar^2$ and a singlet $\ket{\Phi}$ with zero total spin. 
Finally we note that, because $\left\{\hat{s}_i\right\}_{i=1}^{3}$ for the two-particle state obey the canonical algebra (\ref{[s^i,s^j]}), the corresponding spin uncertainty relations are given by Eqs. (\ref{[s^i,s^j]_UR}).

It is straightforward to construct the  two-particle creation and annihilation operators,
\begin{eqnarray} \label{}
\hat{s}_{\pm} = \hat{s}_{\rm x} \pm i \hat{s}_{\rm y} = \hat{s}_{\pm}^{\rm (A)} + \hat{s}_{\pm}^{\rm (B)} \, . 
\end{eqnarray} 
Together with the total $z$-spin of the two-particle state, $\hat{s}_{\rm z} = \hat{s}_{\rm z}^{\rm (A)} + \hat{s}_{\rm z}^{\rm (B)}$, these satisfy the algebra (\ref{s_pm_Lie_algebra}).

Written explicitly, $\hat{s}_{\pm}$ for the two-particle state take the form
\begin{eqnarray} \label{s_pm_explicit*}
\hat{s}_{+} 
= \hbar
\begin{bmatrix}
    0  & 1 & 1 & 0 \\
    0  & 0 & 0 & 1 \\  
    0  & 0 & 0 & 1 \\
    0  & 0 & 0 & 0 \,
\end{bmatrix}
\, , \quad
\hat{s}_{-} 
= \hbar
\begin{bmatrix}
    0  & 0 & 0 & 0 \\
    1  & 0 & 0 & 0 \\  
    1  & 0 & 0 & 0 \\
    0  & 1 & 1 & 0 \,
\end{bmatrix}
\, .
\end{eqnarray}
The eigenvectors of $\hat{s}_{+}$ are $\ket{\Psi_1} = \ket{\uparrow_{\rm z}}_{\rm A}\ket{\uparrow_{\rm z}}_{\rm B} = (1,0,0,0)$ (\ref{s_i^(AB)_eigenvector-1}) and $\ket{\Phi} = 2^{-1/2}(\ket{\uparrow_{\rm z}}_{\rm A}\ket{\downarrow_{\rm z}}_{\rm B} - \ket{\downarrow_{\rm z}}_{\rm A}\ket{\uparrow_{\rm z}}_{\rm B}) = 2^{-1/2}(0,1,-1,0)$ (\ref{s_i^(AB)_eigenvector-4}), plus two copies of the null vector, $|\rm null \rangle = (0,0,0,0)$. 
Three of the eigenvectors of $\hat{s}_{-}$ are the same, but $\ket{\Psi_1}$ is replaced by $\ket{\Psi_2} = \ket{\downarrow_{\rm z}}_{\rm A}\ket{\downarrow_{\rm z}}_{\rm B} = (0,0,0,1)$ (\ref{s_i^(AB)_eigenvector-2}). 
As in the one-particle case, every eigenvector of $\hat{s}_{+}$ and $\hat{s}_{-}$ corresponds to the eigenvalue $0$. 
The two-particle creation and annihilation operators perform spin flips according to:
\begin{subequations} \label{spin_flip_z*} 
\begin{eqnarray} \label{spin_flip_z-*} 
\hat{s}_{-} \ket{\uparrow_{\rm z}}_{\rm A}\ket{\uparrow_{\rm z}}_{\rm B} &=& \hbar \, (\ket{\uparrow_{\rm z}}_{\rm A}\ket{\downarrow_{\rm z}}_{\rm B} + \ket{\downarrow_{\rm z}}_{\rm A}\ket{\uparrow_{\rm z}}_{\rm B})  \, ,  
\nonumber\\
\hat{s}_{-} \ket{\uparrow_{\rm z}}_{\rm A}\ket{\downarrow_{\rm z}}_{\rm B} &=& \hat{s}_{-} \ket{\downarrow_{\rm z}}_{\rm A}\ket{\uparrow_{\rm z}}_{\rm B} = \hbar \, \ket{\downarrow_{\rm z}}_{\rm A}\ket{\downarrow_{\rm z}}_{\rm B} \, ,
\nonumber\\
\hat{s}_{-} \ket{\downarrow_{\rm z}}_{\rm A}\ket{\downarrow_{\rm z}}_{\rm B} &=& 0 \, ,
\end{eqnarray}
\begin{eqnarray} \label{spin_flip_z+*} 
\hat{s}_{+} \ket{\uparrow_{\rm z}}_{\rm A}\ket{\uparrow_{\rm z}}_{\rm B} &=& 0 \, , 
\nonumber\\
\hat{s}_{+} \ket{\uparrow_{\rm z}}_{\rm A}\ket{\downarrow_{\rm z}}_{\rm B} &=& \hat{s}_{+} \ket{\downarrow_{\rm z}}_{\rm A}\ket{\uparrow_{\rm z}}_{\rm B} = \hbar \, \ket{\uparrow_{\rm z}}_{\rm A}\ket{\uparrow_{\rm z}}_{\rm B} \, ,
\nonumber\\
\hat{s}_{+} \ket{\downarrow_{\rm z}}_{\rm A}\ket{\downarrow_{\rm z}}_{\rm B} &=&  \hbar \, (\ket{\uparrow_{\rm z}}_{\rm A}\ket{\downarrow_{\rm z}}_{\rm B} + \ket{\downarrow_{\rm z}}_{\rm A}\ket{\uparrow_{\rm z}}_{\rm B}) \, .
\end{eqnarray}
\end{subequations}

%%%%%%%%%%%%%%%%%%%%%%%%%%%%%%%%%%%%%%%%%%%%%%%%%%%%%%%%%%
\subsection{Bell states} \label{Sec.A.3}

The Bell states for spins in the $i^{\rm th}$ direction are constructed as
\begin{subequations} \label{Bell_states}
\begin{eqnarray} \label{Psi_pm}
\ket{\Psi_{\pm}^{(i)}} = \frac{1}{\sqrt{2}}(\ket{\uparrow_{i}}_{\rm A}\ket{\downarrow_{i}}_{\rm B} \pm \ket{\downarrow_{i}}_{\rm A}\ket{\uparrow_{i}}_{\rm B}) \, , 
\end{eqnarray}
\begin{eqnarray} \label{Phi_pm}
\ket{\Phi_{\pm}^{(i)}} = \frac{1}{\sqrt{2}}(\ket{\uparrow_{i}}_{\rm A}\ket{\uparrow_{i}}_{\rm B} \pm \ket{\downarrow_{i}}_{\rm A}\ket{\downarrow_{i}}_{\rm B}) \, . 
\end{eqnarray}
\end{subequations}
These may be rewritten in terms of the simultaneous eigenvectors of $\hat{s}_{i}$ and $\hat{s}^2$, giving
\begin{subequations} \label{Bell_states*}
\begin{eqnarray} \label{Psi_+}
\ket{\Psi_{+}^{(i)}} = \ket{\Psi_3^{(i)}} = \ket{\hat{s}_{i}=0,\hat{s}^2=2\hbar^2} \, , 
\end{eqnarray}
\begin{eqnarray} \label{Psi_-}
\ket{\Psi_{-}^{(i)}} &=& \ket{\Phi^{(i)}} = \ket{\hat{s}_{i}=0,\hat{s}^2=0} \, , 
\end{eqnarray}
\begin{eqnarray} \label{Phi_+}
\ket{\Phi_{+}^{(i)}} &=& \frac{1}{\sqrt{2}}(\ket{\Psi_1^{(i)}}+\ket{\Psi_2^{(i)}}) 
\nonumber\\ 
&=&  \frac{1}{\sqrt{2}}(\ket{\hat{s}_{i}=+\hbar,\hat{s}^2=2\hbar^2} + \ket{\hat{s}_{i}=-\hbar,\hat{s}^2=2\hbar^2}) \, , 
\nonumber\\ 
\end{eqnarray}
\begin{eqnarray} \label{Phi_-}
\ket{\Phi_{-}^{(i)}} &=& \frac{1}{\sqrt{2}}(\ket{\Psi_1^{(i)}}-\ket{\Psi_2^{(i)}}) 
\nonumber\\ 
&=&  \frac{1}{\sqrt{2}}(\ket{\hat{s}_{i}=+\hbar,\hat{s}^2=2\hbar^2} - \ket{\hat{s}_{i}=-\hbar,\hat{s}^2=2\hbar^2}) \, , 
\nonumber\\ 
\end{eqnarray}
\end{subequations}
where $\left\{\ket{\Psi_1^{(i)}},\ket{\Psi_2^{(i)}},\ket{\Psi_3^{(i)}},\ket{\Phi^{(i)}}\right\}$ are defined, for any $i \in \left\{\rm x,y,z\right\}$, by analogy with (\ref{s_i^(AB)_eigenvector-1})-(\ref{s_i^(AB)_eigenvector-4}). 

However, by convention, $\ket{\Psi_{\pm}}$ and $\ket{\Phi_{\pm}}$ are usually defined with respect to the $z$-axis, so that the superscript denoting direction may be dispensed with. 
Using Eqs. (\ref{y_up_down})-(\ref{x_up_down}) it is straightforward to show that, for any group of Bell states $\left\{\ket{\Psi_{\pm}^{(i)}},\ket{\Phi_{\pm}^{(i)}}\right\}$, the particle spins are entangled with respect to measurements along any axis. 

%%%%%%%%%%%%%%%%%%%%%%%%%%%%%%%%%%%%%%%%%%%%%%%%%%%%%%%%%%
%%%%%%%%%%%%%%%%%%%%%%%%%%%%%%%%%%%%%%%%%%%%%%%%%%%%%%%%%%
\section{The explicit forms of $\hat{{\rm S}}_{\rm z}$, $\hat{{\rm S}}^2$ and $\hat{{\rm S}}_{\rm \pm}$ for the smeared two-particle state} \label{Appendix-B}

Written explicitly, $\hat{{\rm S}}_{\rm z}$ and $\hat{{\rm S}}^2$ for the two-particle state in smeared space take the form
%\begin{widetext}
\begin{eqnarray} \label{S^2_AB_explicit}
\begin{split}
&&\hat{{\rm S}}_{\rm z} = (\hbar+\beta) (1+\delta)^{-1} \times
\nonumber\\ 
&&\left[\begin{matrix} 
    1+\delta & 0 & 0 & 0 & 0 & 0 & 0 & 0 \\% & 0 & 0 & 0 & 0 & 0 & 0 & 0 & 0 \\
    0  & 1 & i\sqrt{\delta} & 0 & 0 & 0 & 0 & 0 \\% & 0 & 0 & 0 & 0 & 0 & 0 & 0 & 0 \\  
    0  & -i\sqrt{\delta} & \delta & 0 & 0 & 0 & 0 & 0 \\% & 0 & 0 & 0 & 0 & 0 & 0 & 0 & 0 \\
    0  & 0 & 0 & 0 & 0 & 0 & 0 & 0 \\% & 0 & 0 & 0 & 0 & 0 & 0 & 0 & 0 \\
    0  & 0 & 0 & 0 & 1 & 0 & 0 & 0 \\% & i\sqrt{\delta} & 0 & 0 & 0 & 0 & 0 & 0 & 0 \\
    0  & 0 & 0 & 0 & 0 & 1-\delta & i\sqrt{\delta} & 0 \\% & 0 & i\sqrt{\delta} & 0 & 0 & 0 & 0 & 0 & 0 \\
    0  & 0 & 0 & 0 & 0 & -i\sqrt{\delta} & 0 & 0 \\% & 0 & 0 & i\sqrt{\delta} & 0 & 0 & 0 & 0 & 0 \\
    0  & 0 & 0 & 0 & 0 & 0 & 0 & -\delta \\% & 0 & 0 & 0 & i\sqrt{\delta} & 0 & 0 & 0 & 0 \\
    0  & 0 & 0 & 0 & -i\sqrt{\delta} & 0 & 0 & 0 \\% & \delta & 0 & 0 & 0 & 0 & 0 & 0 & 0 \\
    0  & 0 & 0 & 0 & 0 & -i\sqrt{\delta} & 0 & 0 \\% & 0 & 0 & i\sqrt{\delta} & 0 & 0 & 0 & 0 & 0 \\
    0  & 0 & 0 & 0 & 0 & 0 & -i\sqrt{\delta} & 0 \\% & 0 & -i\sqrt{\delta} & -(1-\delta) & 0 & 0 & 0 & 0 & 0 \\
    0  & 0 & 0 & 0 & 0 & 0 & 0 & -i\sqrt{\delta} \\% & 0 & 0 & 0 & -1 & 0 & 0 & 0 & 0 \\
    0  & 0 & 0 & 0 & 0 & 0 & 0 & 0 \\% & 0 & 0 & 0 & 0 & 0 & 0 & 0 & 0 \\
    0  & 0 & 0 & 0 & 0 & 0 & 0 & 0 \\% & 0 & 0 & 0 & 0 & 0 & -\delta & i\sqrt{\delta} & 0 \\
    0  & 0 & 0 & 0 & 0 & 0 & 0 & 0 \\% & 0 & 0 & 0 & 0 & 0 & -i\sqrt{\delta} & -1 & 0 \\
    0  & 0 & 0 & 0 & 0 & 0 & 0 & 0 \\% & 0 & 0 & 0 & 0 & 0 & 0 & 0 & -(1+\delta) \,
\end{matrix} 
\right. 
\nonumber\\
&&\left.
\begin{matrix}
    0 & 0 & 0 & 0 & 0 & 0 & 0 & 0 \\
    0 & 0 & 0 & 0 & 0 & 0 & 0 & 0 \\  
    0 & 0 & 0 & 0 & 0 & 0 & 0 & 0 \\
    0 & 0 & 0 & 0 & 0 & 0 & 0 & 0 \\
    i\sqrt{\delta} & 0 & 0 & 0 & 0 & 0 & 0 & 0 \\
    0 & i\sqrt{\delta} & 0 & 0 & 0 & 0 & 0 & 0 \\
    0 & 0 & i\sqrt{\delta} & 0 & 0 & 0 & 0 & 0 \\
    0 & 0 & 0 & i\sqrt{\delta} & 0 & 0 & 0 & 0 \\
    \delta & 0 & 0 & 0 & 0 & 0 & 0 & 0 \\
    0 & 0 & i\sqrt{\delta} & 0 & 0 & 0 & 0 & 0 \\
    0 & -i\sqrt{\delta} & -(1-\delta) & 0 & 0 & 0 & 0 & 0 \\
    0 & 0 & 0 & -1 & 0 & 0 & 0 & 0 \\
    0 & 0 & 0 & 0 & 0 & 0 & 0 & 0 \\
    0 & 0 & 0 & 0 & 0 & -\delta & i\sqrt{\delta} & 0 \\
    0 & 0 & 0 & 0 & 0 & -i\sqrt{\delta} & -1 & 0 \\
    0 & 0 & 0 & 0 & 0 & 0 & 0 & -(1+\delta) \,
\end{matrix}\right]
\end{split} 
\nonumber\\ 
\end{eqnarray}
%\end{widetext}
%
%\begin{widetext}
\begin{eqnarray} \label{S_z_AB_explicit}
\begin{split}
&&\hat{{\rm S}}^{2} = (\hbar+\beta)^2 (1+\delta)^{-1} \times
\nonumber\\
&&\left[\begin{matrix}
2(1+\delta) & 0 & 0 & 0 & 0 & 0 & 0 & 0 \\ %
 0 & 2+\delta & i\sqrt{\delta} & 0 & \delta & 0 & 0 & 0 \\ %
 0 & -i\sqrt{\delta} & 1 + 2\delta & 0 & i\sqrt{\delta} & 0 & 0 & 0 \\ %
 0 & 0 & 0 & 1+\delta & 0 & \frac{2\delta +i(1-\delta)\sqrt{\delta}}{1+\delta} & -\frac{(i\delta +\sqrt{\delta})^2}{1+\delta} & 0 \\ %
 0 & \delta & -i\sqrt{\delta} & 0 & 2 + \delta & 0 & 0 & 0 \\ %
 0 & 0 & 0 & \frac{2\delta - i(1-\delta)\sqrt{\delta}}{1+\delta} & 0 & \frac{2(1+\delta+\delta^2)}{1+\delta} & \frac{i(1-\delta)\sqrt{\delta}}{1+\delta} & 0 \\ %
 0 & 0 & 0 & -\frac{(i\delta + \sqrt{\delta})^2}{1+\delta} & 0 & -\frac{i(1 -\delta)\sqrt{\delta}}{1+\delta} & \frac{1+4\delta + \delta^2}{1+\delta} & 0 \\ %
 0 & 0 & 0 & 0 & 0 & 0 & 0 & 1 + 2\delta \\ %
 0 & i\sqrt{\delta} & 1 & 0 & -i\sqrt{\delta} & 0 & 0 & 0 \\ %
 0 & 0 & 0 & -\frac{(i+\sqrt{\delta})^2}{1+\delta} & 0 & -\frac{i(1-\delta)\sqrt{\delta}}{1+\delta} & \frac{2\delta}{1+\delta} & 0 \\ %
 0 & 0 & 0 & \frac{2\delta - i(1-\delta)\sqrt{\delta}}{1+\delta} & 0 & -\frac{2\delta}{1+\delta} & \frac{i(1-\delta)\sqrt{\delta}}{1+ \delta} & 0 \\ %
 0 & 0 & 0 & 0 & 0 & 0 & 0 & i\sqrt{\delta} \\ %
 0 & 0 & 0 & 0 & 0 & \frac{2\delta + i(1-\delta)\sqrt{\delta}}{1+\delta} & -\frac{(i + \sqrt{\delta})^2}{1+\delta} & 0 \\ %
 0 & 0 & 0 & 0 & 0 & 0 & 0 \\ %
 0 & 0 & 0 & 0 & 0 & 0 & 0 & -i\sqrt{\delta} \\ %
 0 & 0 & 0 & 0 & 0 & 0 & 0 & 0 \, 
\end{matrix}
\right. 
\nonumber\\ 
&&\left.
\begin{matrix}
 0 & 0 & 0 & 0 & 0 & 0 & 0 & 0 \\ %
 -i\sqrt{\delta} & 0 & 0 & 0 & 0 & 0 & 0 & 0 \\ %
 1 & 0 & 0 & 0 & 0 & 0 & 0 & 0 \\ %
 0 & -\frac{(i + \sqrt{\delta})^2}{1+\delta} & \frac{2\delta +i(1- \delta)\sqrt{\delta}}{1+\delta} & 0 & 0 & 0 & 0 & 0 \\ %
 i\sqrt{\delta} & 0 & 0 & 0 & 0 & 0 & 0 & 0 \\ %
 0 & \frac{i(1 - \delta)\sqrt{\delta}}{1+\delta} & -\frac{2\delta}{1+\delta} & 0 & \frac{2\delta - i(1-\delta)\sqrt{\delta}}{1+\delta} & 0 & 0 & 0 \\ %
 0 & \frac{2\delta}{1+\delta} & -\frac{i(1 - \delta)\sqrt{\delta}}{1+\delta} & 0 & -\frac{(i+\sqrt{\delta})^2}{1+\delta} & 0 & 0 & 0 \\ %
 0 & 0 & 0 & -i\sqrt{\delta} & 0 & 1 & i\sqrt{\delta} & 0 \\ %
 1 + 2\delta & 0 & 0 & 0 & 0 & 0 & 0 & 0 \\ %
 0 & \frac{1+4\delta+\delta^2)}{1+\delta} & -\frac{i(1-\delta)\sqrt{\delta}}{1+\delta} & 0 & -\frac{(i\delta + \sqrt{\delta})^2}{1+\delta} & 0 & 0 & 0 \\ %
 0 & \frac{i(1-\delta)\sqrt{\delta}}{1+ \delta} & \frac{2(1+\delta+\delta^2)}{1+\delta} & 0 & \frac{2\delta - i(1-\delta)\sqrt{\delta}}{1+\delta} & 0 & 0 & 0 \\ %
 0 & 0 & 0 & 2 + \delta & 0 & -i\sqrt{\delta} & \delta & 0 \\ %
 0 & -\frac{(-i\delta+\sqrt{\delta})^2}{1+\delta} & \frac{2\delta + i(1-\delta)\sqrt{\delta}}{1+\delta} & 0 & 1+\delta & 0 & 0 & 0 \\ %
 0 & 0 & 0 & i\sqrt{\delta} & 0 & 1+2\delta & -i\sqrt{\delta} & 0 \\ %
 0 & 0 & 0 & \delta & 0 & i\sqrt{\delta} & 2 + \delta & 0 \\ %
 0 & 0 & 0 & 0 & 0 & 0 & 0 & 2(1+\delta) \, 
\end{matrix}\right]
\end{split} 
\nonumber\\ 
\end{eqnarray}
%\end{widetext}
and the corresponding creation and annihilation operators are:
%\begin{widetext}
\begin{eqnarray} \label{S_plus_AB_explicit}
\begin{split}
&&\hat{{\rm S}}_{+} = (\hbar+\beta) (1+\delta)^{-1} \times
\nonumber\\ 
&&\left[\begin{matrix} 
 0 & \delta + i\sqrt{\delta} & 1 - i\sqrt{\delta} & 0 & \delta + i\sqrt{\delta} & 0 & 0 & 0 \\ %
 0 & 0 & 0 & 1 + i\sqrt{\delta} & 0 & \delta + i\sqrt{\delta} & 0 & 0 \\ %
 0 & 0 & 0 & \delta - i\sqrt{\delta} & 0 & 0 & \delta + i\sqrt{\delta} & 0 \\ %
 0 & 0 & 0 & 0 & 0 & 0 & 0 & \delta + i\sqrt{\delta} \\ %
 0 & 0 & 0 & 0 & 0 & \delta + i\sqrt{\delta} & 1 - i\sqrt{\delta} & 0 \\ %
 0 & 0 & 0 & 0 & 0 & 0 & 0 & 1 + i\sqrt{\delta} \\ %
 0 & 0 & 0 & 0 & 0 & 0 & 0 & \delta - i\sqrt{\delta} \\ %
 0 & 0 & 0 & 0 & 0 & 0 & 0 & 0 \\ %
 0 & 0 & 0 & 0 & 0 & 0 & 0 & 0 \\ %
 0 & 0 & 0 & 0 & 0 & 0 & 0 & 0 \\ %
 0 & 0 & 0 & 0 & 0 & 0 & 0 & 0 \\ %
 0 & 0 & 0 & 0 & 0 & 0 & 0 & 0 \\ %
 0 & 0 & 0 & 0 & 0 & 0 & 0 & 0 \\ %
 0 & 0 & 0 & 0 & 0 & 0 & 0 & 0 \\ %
 0 & 0 & 0 & 0 & 0 & 0 & 0 & 0 \\ %
 0 & 0 & 0 & 0 & 0 & 0 & 0 & 0 \, %   
\end{matrix} \, , 
\right. 
\nonumber\\ 
&&\left.
\begin{matrix} 
 1 - i\sqrt{\delta} & 0 & 0 & 0 & 0 & 0 & 0 & 0 \\ %
 0 & 1 - i\sqrt{\delta} & 0 & 0 & 0 & 0 & 0 & 0 \\ %
 0 & 0 & 1 - i\sqrt{\delta} & 0 & 0 & 0 & 0 & 0 \\ %
 0 & 0 & 0 & 1 - i\sqrt{\delta} & 0 & 0 & 0 & 0 \\ %
 0 & 0 & 0 & 0 & 1 + i\sqrt{\delta} & 0 & 0 & 0 \\ %
 0 & 0 & 0 & 0 & 0 & 1 + i\sqrt{\delta} & 0 & 0 \\ %
 0 & 0 & 0 & 0 & 0 & 0 & 1 + i\sqrt{\delta} & 0 \\ %
 0 & 0 & 0 & 0 & 0 & 0 & 0 & 1 + i\sqrt{\delta} \\ %
 0 & \delta + i\sqrt{\delta} & 1 - i\sqrt{\delta} & 0 & \delta - i\sqrt{\delta} & 0 & 0 & 0 \\ %
 0 & 0 & 0 & 1 + i\sqrt{\delta} & 0 & \delta - i\sqrt{\delta} & 0 & 0 \\ %
 0 & 0 & 0 & \delta - i\sqrt{\delta} & 0 & 0 & \delta - i\sqrt{\delta} & 0 \\ %
 0 & 0 & 0 & 0 & 0 & 0 & 0 & \delta - i\sqrt{\delta} \\ %
 0 & 0 & 0 & 0 & 0 & \delta + i\sqrt{\delta} & 1 - i\sqrt{\delta} & 0 \\ %
 0 & 0 & 0 & 0 & 0 & 0 & 0 & 1 + i\sqrt{\delta} \\ %
 0 & 0 & 0 & 0 & 0 & 0 & 0 & \delta - i\sqrt{\delta} \\ %
 0 & 0 & 0 & 0 & 0 & 0 & 0 & 0 \, % 
\end{matrix}\right] \, , 
\end{split}
\nonumber\\
\end{eqnarray}
%\end{widetext}
%
%\begin{widetext}
\begin{eqnarray} \label{S_minus_AB_explicit}
\begin{split}
&&\hat{{\rm S}}_{-} = (\hbar+\beta) (1+\delta)^{-1} \times
\nonumber\\ 
&&\left[\begin{matrix} 
 0 & 0 & 0 & 0 & 0 & 0 & 0 & 0 \\ %
 \delta - i\sqrt{\delta} & 0 & 0 & 0 & 0 & 0 & 0 & 0 \\ %
 1 + i\sqrt{\delta} & 0 & 0 & 0 & 0 & 0 & 0 & 0 \\ %
 0 & 1 - i\sqrt{\delta} & \delta + i\sqrt{\delta} & 0 & 0 & 0 & 0 & 0 \\ %
 \delta - i\sqrt{\delta} & 0 & 0 & 0 & 0 & 0 & 0 & 0 \\ %
 0 & \delta - i\sqrt{\delta} & 0 & 0 & \delta - i\sqrt{\delta} & 0 & 0 & 0 \\ %
 0 & 0 & \delta - i\sqrt{\delta} & 0 & 1 + i\sqrt{\delta} & 0 & 0 & 0 \\ %
 0 & 0 & 0 & \delta - i\sqrt{\delta} & 0 & 1 - i\sqrt{\delta} & \delta + i\sqrt{\delta} & 0 \\ %
 1 + i\sqrt{\delta} & 0 & 0 & 0 & 0 & 0 & 0 & 0 \\ %
 0 & 1 + i\sqrt{\delta} & 0 & 0 & 0 & 0 & 0 & 0 \\ %
 0 & 0 & 1 + i\sqrt{\delta} & 0 & 0 & 0 & 0 & 0 \\ %
 0 & 0 & 0 & 1 + i\sqrt{\delta} & 0 & 0 & 0 & 0 \\ %
 0 & 0 & 0 & 0 & 1 - i\sqrt{\delta} & 0 & 0 & 0 \\ %
 0 & 0 & 0 & 0 & 0 & 1 - i\sqrt{\delta} & 0 & 0 \\ %
 0 & 0 & 0 & 0 & 0 & 0 & 1 - i\sqrt{\delta} & 0 \\ %
 0 & 0 & 0 & 0 & 0 & 0 & 0 & 1 - i\sqrt{\delta} \, %
\end{matrix} \, , 
\right. 
\nonumber\\ 
&&\left.
\begin{matrix} 
0 & 0 & 0 & 0 & 0 & 0 & 0 & 0 \\ %
0 & 0 & 0 & 0 & 0 & 0 & 0 & 0 \\ %
0 & 0 & 0 & 0 & 0 & 0 & 0 & 0 \\ %
0 & 0 & 0 & 0 & 0 & 0 & 0 & 0 \\ %
0 & 0 & 0 & 0 & 0 & 0 & 0 & 0 \\ %
0 & 0 & 0 & 0 & 0 & 0 & 0 & 0 \\ %
0 & 0 & 0 & 0 & 0 & 0 & 0 & 0 \\ %
0 & 0 & 0 & 0 & 0 & 0 & 0 & 0 \\ %
0 & 0 & 0 & 0 & 0 & 0 & 0 & 0 \\ %
\delta - i\sqrt{\delta} & 0 & 0 & 0 & 0 & 0 & 0 & 0 \\ %
1 + i\sqrt{\delta} & 0 & 0 & 0 & 0 & 0 & 0 & 0 \\ %
0 & 1 - i\sqrt{\delta} & \delta + i\sqrt{\delta} & 0 & 0 & 0 & 0 & 0 \\ %
\delta + i\sqrt{\delta} & 0 & 0 & 0 & 0 & 0 & 0 & 0 \\ %
0 & \delta + i\sqrt{\delta} & 0 & 0 & \delta - i\sqrt{\delta} & 0 & 0 & 0 \\ %
0 & 0 & \delta + i\sqrt{\delta} & 0 & 1 + i\sqrt{\delta} & 0 & 0 & 0 \\ %
0 & 0 & 0 & \delta + i\sqrt{\delta} & 0 & 1 - i\sqrt{\delta} & \delta + i\sqrt{\delta} & 0 \, %
\end{matrix}\right] \, .
\end{split}
\nonumber\\
\end{eqnarray}
%\end{widetext}

%Bilbiography%%%%%%%%%%%%%%%%%%%%%%%%%%%%%%%%%%%%%%%%%%%%%%%%%%%
%%%%%%%%%%%%%%%%%%%%%%%%%%%%%%%%%%%%%%%%%%%%%%%%%%%%%%%%%%

\end{document}